\documentstyle[here,epsfig]{mn2e}
\def\msun{{\rm{M}}_{\odot}}
\def\simlt{\mathrel{\rlap{\lower 3pt\hbox{$\sim$}}\raise 2.0pt\hbox{$<$}}}
\def\simgt{\mathrel{\rlap{\lower 3pt\hbox{$\sim$}} \raise 2.0pt\hbox{$>$}}}
\def\lsim{\mathrel{\rlap{\lower 3pt\hbox{$\sim$}}\raise 2.0pt\hbox{$<$}}}
\def\gsim{\mathrel{\rlap{\lower 3pt\hbox{$\sim$}} \raise 2.0pt\hbox{$>$}}}

\begin{document}

\title{Millisecond pulsars around intermediate--mass black holes in
globular clusters}
\author[B. Devecchi, M. Colpi, M. Mapelli, A. Possenti]{B.~
Devecchi$^1$, M.~Colpi$^1$, M.~Mapelli$^2$, \& A.~Possenti$^3$\\ 
$1$ Dipartimento di Fisica G.~Occhialini, Universit\`a degli Studi di
Milano Bicocca, Piazza della Scienza 3, I-20126 Milano, Italy\\ 
$2$ Institute for Theoretical Physics, University of Z$\ddot {\rm u}$rich, 
Winterthurerstrasse 190, CH-8057 Zürich, Switzerland \\
$3$ INAF, Osservatorio Astronomico di Cagliari, Poggio dei Pini, Strada 54,
I-09012 Capoterra, Italy}

\maketitle \vspace {0.2cm}

\begin{abstract}
  Globular clusters (GCs) are expected to be breeding grounds for the
  formation of single or binary intermediate--mass black holes (IMBHs)
  of $\simgt 100\,\msun,$ but a clear signature of their existence is
  still missing. In this context, we study the process of dynamical
  capture of a millisecond pulsar (MSP) by a single or binary IMBH,
  simulating various types of single-binary and binary-binary
  encounters.  It is found that [IMBH,MSP] binaries form over cosmic
  time in a cluster, at rates $\lsim 10^{-11}$ ${\rm yr}^{-1}$, via
  encounters of wide--orbit binary MSPs off the single IMBH, and at a
  lower pace, via interactions of (binary or single) MSPs with the
  IMBH orbited by a typical cluster star.  The formation of an
  [IMBH,MSP] system is strongly inhibited if the IMBH is orbited by a
  stellar mass black hole (BH): in this case, the only viable path is
  through the formation of a rare stable hierarchical triplet with the
  MSP orbiting exterior to the [IMBH,BH] binary.  The [IMBH,MSP]
  binaries that form are relatively short-lived, $\lsim 10^{8-9}$ yr,
  since their orbits decay via emission of gravitational waves.  The
  detection of an [IMBH,MSP] system has a low probability of occurrence, 
  when inferred from the current sample of MSPs in
  GCs.  If next generation radio telescopes, like SKA, will detect an
  order of magnitude larger population of MSP in GCs, at least one
  [IMBH,MSP] is expected.  Therefore, a complete search for
  low-luminosity MSPs in the GCs of the Milky Way with SKA will have
  the potential of testing the hypothesis that IMBHs of order
  $100\,\msun$ are commonly hosted in GCs.  The discovery will
  unambiguously prove that black holes exist in the still uncharted
  interval of masses around $\simgt 100\,\msun$.
\end{abstract}


\section{Introduction}

\subsection{IMBHs: Observations}

A number of observations suggest that intermediate--mass black holes
(IMBHs) may exist with masses between $\approx 100\,\msun$ to
$10^4\,\msun$. Heavier than the stellar-mass black holes (BHs) born in
core-collapse supernovae ($3\,\msun-30\,\msun$; Orosz 2003), IMBHs are
expected to form in dense, rich stellar systems through complex
dynamical processes.  Globular clusters (GCs), among the densest
stellar systems known in galaxies, have therefore become prime sites
for their search.

Gebhardt, Rich \& Ho (2002, 2005) suggested the presence of an IMBH of
$2^{+1.4}_{-0.8}\times 10^4\,\msun,$ in the cluster G1 of M31, on the
basis of a joined analysis of photometric and spectroscopic
measurements.  Remarkably, the IMBH in G1 seems to lie just on the
low--end of the BH mass versus one--dimensional dispersion velocity
correlation observed in spheroids and bulges of nearby galaxies
(Ferrarese \& Merritt 2000; Gebhardt et al. 2000).

In the galactic GC M15, {\it HST} and ground--based observations of
line--of--sight velocities and proper motions, indicated the
occurrence of a central concentration of non--luminous matter of
$500^{+2500}_{-500}\, \msun$, that could be ascribed to the presence
of an IMBH (van den Bosch et al. 2006; Gerssen et al. 2002).  By
mapping the velocity field, van den Bosch et al. (2006) found also
evidence of ordered rotation in the central 4 arc sec of M15. This
unexpected dynamical state in a region of rapid relaxation ($10^7$ yr)
may give first evidence, albeit indirect, that a source of angular
momentum in the form of a``binary'' IMBH may exist in M15 (Mapelli et
al. 2005).  Claims of the possible presence of an IMBH have been
advanced also in 47 Tucanae (McLaughlin et al. 2006).

An additional puzzling picture has emerged from observations in the GC
NGC 6752. Two millisecond pulsars (MSPs hereon), PSR-B and PSR-E, show
unusual accelerations (D'Amico et al. 2002), that, once ascribed to
the overall effect of the cluster potential well, indicate the
presence of $\simgt 1000\, \msun$ of under--luminous matter enclosed
within the central 0.08 pc (Ferraro et al. 2003a).  NGC 6752 is even
more peculiar than M15, since it also hosts two MSPs with unusual
locations. PSR-A, a binary pulsar with a white dwarf (WD) companion
(D'Amico et al. 2002; Ferraro et al. 2003b; Bassa et al. 2003) and a
very low orbital eccentricity ($\sim 10^{-5}$, D'Amico et al. 2002)
holds the record of being the farthest MSP ever observed in a GC, at a
distance of $\approx 3.3$ half mass radii. PSR-C, an isolated MSP,
ranks second in the list of the most offset pulsars known, at a
distance of 1.4 half mass radii from the gravitational center of the
cluster (D'Amico et al. 2002; Corongiu et al. 2006).  Colpi, Possenti
\& Gualandris (2002) first conjectured that PSR-A was propelled into
the halo in a flyby off a binary BH in the mass range between
$10\,\msun$ and $100\,\msun$ opening the perspective of unveiling
binary BHs in GCs (see Section $ 1.2$).  Prompted by the evidence of
under-luminous matter in the core of NGC 6752, Colpi, Mapelli \&
Possenti (2003) carried on an extensive analysis of binary-binary
encounters with IMBHs, to asses the viability of this scenario.  They
found that a $\sim 100\,\msun$ IMBH with a stellar--mass BH in a
binary would be the best target for imprinting the necessary thrust to
PSR-A\footnote {Ejection of PSR-A from the core to the halo following
  exchange interactions off normal binary stars can not be excluded,
  but as pointed out by Colpi et al. (2002; Sigurdsson 2003), the
  binary parameters of PSR-A and its evolution make this possibility
  remote, and call for fine tuning conditions.} and at the same time
for preserving the low eccentricity of the binary pulsar (within a
factor of 3 for the bulk of the simulated encounters). Instead, larger
mass IMBHs ($\sim{} 500\,{}M_\odot{}$) with star companions can
produce the correct ejection velocity, but cause the eccentricity to
grow much larger. Thus, PSR-A had to interact with the very massive
IMBH only before its recycling phase.

The observation of IMBHs in GCs is still far from being conclusive,
since numerical studies have shown that kinematic features as those
observed in G1 and M15 can be reproduced assuming, in the cluster
center, the presence of a collection of low--mass compact remnants,
with no need of a single massive IMBH (Baumgardt et al. 2003a,b).  In
addition, a single massive ($\gsim 1000\, \msun$) IMBH, if present,
would affect the stellar dynamics (because of energy generation in the
IMBH cusp) creating a constant density profile of bright stars in
projection that differs from the typical profile of a core-collapse
cluster such as M15 (Baumgardt, Makino \& Ebisuzaki 2004).

\subsection{IMBHs: Theory}

On theoretical ground a number of authors suggested that IMBHs may
form inside either (i) young and dense star clusters vulnerable to
unstable mass segregation and core collapse before the most massive
stars explode as supernovae (Portegies Zwart \& McMillan 2002;
Freitag, Gurkan \& Rasio 2006; G$\ddot {\rm u}$rkan et al. 2006) or
(ii) dynamically in already evolved GCs when all the massive stars
have turned into stellar--mass BHs (Miller \& Hamilton 2002).  In the
first case, runaway collisions among young massive stars may lead to
the formation of a very massive stellar object which ultimately
collapses into an IMBH \footnote{The effects of the environment, of
  rotation and metallicity, on the formation and fate of these
  ultra--massive stars are largely unknown. A recent study on the
  mass loss of merged stars (during and after the merger) of $\sim
  100\,\msun$ have shown that this does not seem to inhibit the
  formation of very massive stars (Suzuki et al. 2007).  However
  further studies are needed in order to better constrain the
  evolution of those more massive object ($\sim 1000\,\msun$) that
  should form $\sim 1000\,\msun$ IMBH.}. In the second case, IMBH
formation requires a succession of close gravitational encounters
among stellar-mass BHs: being the heaviest objects in the cluster,
these BHs may segregate in the core under the action of the Spitzer's
mass stratification instability (Spitzer 1969; Lightman \& Fall 1978;
Watters, Joshi \& Rasio 2000; Khalisi, Amaro-Seoane \& Spurzem 2006),
forming a dense core which becomes dynamically decoupled from the rest
of the stars.  Hardening and recoil among the interacting BHs lead to
their ejection from the cluster (Sigurdsson \& Hernquist 1993;
Kulkarni, Hut \& McMillan 1993; Portegies Zwart \& McMillan 2000) and
at the same time to the increase of their mass because of repeated
mergers (Miller \& Hamilton 2002).  O'Leary et al. (2006) have
recently shown that there is a significant probability (between 20\%
to 80\%) of BH growth, and found final masses $\simgt 100\,\msun$.
After evaporation of most of the BHs on a timescale of $\sim$ Gyr, one
IMBH and/or few BHs, single or in binaries, may remain inside the GC.

The recent discovery of a luminous, highly variable X-ray source in
one GC of NGC 4472 (Maccarone et al. 2007) may have just provided
first evidence that at least one BH is retained inside. Whether this
source in NGC 4472 is an accreting BH or IMBH is still uncertain, but
this finding goes in the direction noted by Pfahl (2005), who
considered the possibility that an IMBH would tidally capture a star
leading to the turn--on of a bright X-ray source.

Given all these uncertainties and the importance of establishing the
possible existence of IMBH in GCs, we explore in this paper an
alternative root, i.e., the possibility that gravitational encounters
off the IMBH provide a path for the dynamical capture of a MSP and the
formation of a binary (hereafter labeled [IMBH,MSP]) comprising the
IMBH and the MSP.  Timing of the radio signal emitted by the MSP would
provide in this way a direct, unambiguous measure of the BH mass.

Motivated by the observation of the halo MSPs in NGC 6752, we simulate
a series of dynamical interactions between a binary MSP and a single
or a binary IMBH, and also between a single MSP and a binary IMBH.  In
the context adopted, the binary IMBH may have a stellar--mass BH, or a
star, as companion.

The outline of the paper is as follows. In Section 2, we describe the
initial conditions of the three and four--body encounters.  In Section
3, we compute cross sections for the formation of [IMBH,MSP] systems
coming from encounters with PSR-A like MSP binaries. We study the
orbital characteristics of the [IMBH,MSP] binaries in their
end-states, and explore the stability of triple systems that form,
against dynamical and resonant self-interactions.  Binary systems
composed by the WD and the IMBH are also considered, and the results
are shortly summarized in Section 4.  In Section 5, we show the
results obtained from simulations with binary MSPs different from
PSR-A that represent the observed population in GCs. We study their
end-states and their characteristic lifetimes taking into account for
their hardening by cluster stars and by gravitational wave driven
in-spiral.  In Section 6 we study the detectability of MSPs around
IMBHs in GCs and discuss the potential importance of these systems for
next-generation deep radio surveys in the Galactic halo. In Section 7
we summarize our findings.

\section{Gravitational encounters}

\begin{table}
\begin{center}
\begin{tabular}[h!]{|ccccccc|}
  \hline
  $M$ ($\msun$) & $a_{\rm{m}}$ (AU) & $a_{\rm{M}}$ (AU) &   N\\
  \hline
  \hline
  100  &- &-  &  5000\\
  \hline
  300  &- &-  & 5000\\
  \hline
  [100,star]  &0.2& 200   & 3000\\
  \hline
  [300,star]  &0.42 &417  & 3000\\
  \hline
$[100,10]_{\rm {h,*}}$ &0.24 &1960   &5000\\
\hline
$[300,10]_{\rm{h,*} }$ &0.4 &5526   &5000\\
\hline
$[100,10]_{\rm gw}$ &$2.2\times10^{-3}$ & 0.24 &10000 \\
\hline
$[300,10]_{\rm gw}$ &$3.2\times10^{-3}$ & 0.4 & 10000\\
\hline
\hline
${\rm [100,star]_{\rm MSP, single}}$  &0.2& 200   & 5000\\
\hline
${\rm [300,star]_{\rm MSP, single}}$  &0.42 &417  & 5000\\
\hline
\end{tabular}
\end{center}
\label{tab:fraz1}
\caption{Initial parameters for simulations with PSR-A like initial MSP binaries. Rows
  refer to different initial states of the IMBH (referred as channels in
  the text). The different columns refer to: selected IMBH mass, minimum
  and maximum values for the distribution of the semi-major axis (for
  the [IMBH,star] and [IMBH,BH] binaries) and
  number of runs for each simulation.  The first eight lines refer to
  encounters with the [MSP,WD] binary, the last two refer to encounters
  with a single MSP.}
\end{table}


\subsection{The projectile}

We consider encounters in which the projectile is either a [MSP,WD]
binary, or a single MSP.  As first case--study, we simulate [MSP,WD]
systems similar to PSR-A in NGC 6752: the MSP has a mass $m_{\rm
  MSP}=1.4\,\msun$ and a WD companion of $m_{\rm WD}=0.2\,\msun$; the
binary has semi-major axis $a_{\rm MSP,i}=0.0223$ AU, orbital period
of 0.86 days, and orbital eccentricity $e_{\rm {MSP,i}}=10^{-5}$. 

We then simulate binary MSPs whose characteristics are extrapolated
from the observed sample of MSPs belonging to the GCs of the Milky Way
(Camilo \& Rasio 2005) (see Section 5 for further discussion).  For
the single MSP, we consider $m_{\rm MSP}=1.4\,\msun$.

\subsection{The target IMBH}

The target is an IMBH, either single or binary, and has no stellar
cusp (Baumgardt et al 2004).  In agreement with O'Leary et al. (2006)
and Colpi et al.(2003), its mass $M_{\rm IMBH}$ is either $100\,\msun$
or $300\,\msun$.

The binary IMBHs have initial semi-major axes and eccentricities drawn
from probability distributions that account for their physical
conditions in a GC.  In details, the initial properties of the target
[IMBH, star] and [IMBH, BH] binaries are the following.

\begin{itemize}
\item[$\bullet$] [IMBH, star]: We randomly generate the mass $m_*$ of the
star, the semi-major axis $a_*$ and the eccentricity $e_*$. The values
for $m_{*}$ follow a current mass function biased toward massive
stars, in order to account for dynamical mass segregation in the core
of the cluster.  We thus consider a mass function $dN/dm\propto
m^{-(1+x)}$ with $x=-5$ as inferred from observations of 47~Tucanae 
(Monkman et al. 2006) with an upper cut--off mass of 0.95 M$_{\odot}.$
For the semi-major axes we follow the analysis proposed by Pfahl
(2005) and briefly summarized in Appendix A.  The values of $a_*$
refer to conditions acquired in dynamical ionization of incoming stellar 
binaries off an initially single IMBH.  Table 1 gives the initial
minimum and maximum semi-major axes used at the onset of the
simulations.  The eccentricity $e_*$ follows a thermal distribution
(Blecha et al. 2006).  The same distribution for $a_*$, $e_*$ and
$m_*$ is used for the interaction of the [IMBH,star] binary
both with [MSP,WD] and single MSP. To distinguish these two cases,
hereon we will refer to the latter using the subscript ``MSP,single''.

\item[$\bullet$] [IMBH, BH]: The IMBH has a BH companion of $m_{\rm
{BH}}=10\,\msun$. The binary has semi-major axis $a_{\rm BH}$ drawn
from two distinct probability distributions, which have been derived:
\noindent
(i) from the hardening due to encounters off cluster stars (subscript
[h,*], hereon), occurring on a time-scale (Quinlan 1996; Mapelli et
al. 2005)
\begin{equation}\label{eq:tharde}
t_{\rm h}(a)\sim \frac{\langle v_*\rangle }{\left(2\pi\xi\right)
G\langle \rho_*\rangle }\frac{1}{a_{\rm BH}}=
2\times 10^7 v_{10}a^{-1}_5 \rho^{-1}_{5.8} {\rm yr,}
\end{equation}
where $\langle \rho_*\rangle $, $\langle v_*\rangle $ and $\xi$ are
the mean stellar mass density, dispersion velocity and hardening
efficiency (we assume $\langle v_*\rangle =10\, v_{10}\,{\rm
  km\,{}s^{-1}}$, $\xi=1$ (Colpi et al. 2003), $a_{\rm BH}=5 a_5
\,{\rm AU}$ and for the density $\langle \rho_*\rangle=7\times
10^5\,\rho_{5.8} \,\msun\,{\rm pc^{-3}}$, the value inferred averaging
over the GC sample currently hosting the population of known MSPs (see
Section 5));

\noindent
(ii) from the in-spiral driven by gravitational wave back--reaction
(subscript [gw], hereon), when the binary is tight (Section 1 of
Appendix A, for details).  The corresponding time-scale, function of
the semi-major axis $a_{\rm BH}$ and eccentricity $e_{\rm BH}$ (Peters
\& Mathews 1963), is:
\begin{eqnarray}\label{eq:tgw}
  t_{\rm gw}(a_{\rm BH},e_{\rm BH})\equiv\frac{5}{256}\frac{c^5a_{\rm
      BH}^4\left(1-e_{\rm BH}^2\right)^{7/2}}{G^3m_{\rm BH}M_{\rm
      IMBH}\left(m_{\rm BH}+M_{\rm IMBH}\right)}\nonumber \\=4.4\times 
  10^8 a^4_{0.2} M^{-1}_{100}m^{-1}_{10} M_{\rm T,110}^{-1} \rm {yr},\nonumber\\
\end{eqnarray}
\noindent
where the following normalizations are used to estimate $t_{\rm gw}$
for $e_{\rm BH}=0.7$: $a_{\rm BH}=0.2 a_{0.2}\,$ AU, $M_{\rm IMBH}=100
M_{100}\,\msun$, $m_{\rm BH}=10 m_{10}\,\msun$, and $M_{\rm T}=M_{\rm
  IMBH}+m_{\rm BH}=110 M_{\rm T,110}\,\msun$.
The peak of the composite semi-major axis distribution occurs when the
two processes become comparable, i.e. at a distance
\begin{equation}\label{eq:agw}
a_{\rm gw}(e_{\rm {BH}}) \sim\left[\frac{256}{5}\frac{G^2 m_{\rm {BH}}
\,M_{\rm {IMBH}} \left(m_{\rm {BH}}+M_{\rm
{IMBH}}\right)\langle v_*\rangle}{\left(1-e_{\rm {BH}}^2\right)^{7/2}c^5
\langle \rho_*\rangle \,{}2\pi\xi}\right]^{1/5}
\end{equation}
corresponding to $t_{\rm h}=t_{\rm gw}$, inferred from equations (1)
and (2).  Typical separations for our [IMBH,BH] binaries are $\sim
0.3$ AU.

\noindent
In the hardening phase by stars the eccentricity $e_{\rm BH}$ is
extracted from a thermal distribution, while during the gravitational
wave driven phase the values of $e_{\rm BH}$ are inferred considering
the modifications induced by gravitational wave loss (see Section 1 of
Appendix A).
\end{itemize}
 
\subsection {Code and outcomes}

We run the numerical code Chain (kindly suited by S. Aarseth) which
makes use of a Bulirsch-Stoer variable step integrator with KS-chain
regularization. The code FEBO (FEw-BOdy), based on a fifth--order
Runge-Kutta scheme (described in Colpi, Mapelli \& Possenti 2003 and
in Mapelli et al. 2005), has been used for trial runs and gives
results in nice agreement with Chain.

The impact parameters of the incoming binaries are distributed
uniformly in $b^2$ (Hut \& Bahcall 1983) up to a maximum value
$b^2_{\rm {max}}$ (see Section 3 of Appendix A).  The phases of the
binaries and the angles describing the initial direction and
inclination of the encounter are extracted from the distributions by
Hut \& Bahcall (1983).  The relative speed $v_{\infty}$ has been
sampled at random from a uniform distribution, in the range 8-12 km
s$^{-1}$, consistent with the values of NGC 6752 (Dubath, Meylan \&
Mayor 1997).  The relative distance between the centers of mass of the
interacting binaries is set equal to the gravitational influence
radius of the target IMBH, $r_{\rm inf}\sim 2GM_{\rm IMBH}/\langle
v_\infty \rangle^2$ ($\sim{}$2000 AU for the $100\,\msun$
case\footnote{For the $300\msun$ IMBH, the larger initial distance
  (6000 AU) makes prohibitive the integration time for the simulations
  run with FEBO. For this reason integration starts at 2000 AU after
  correcting for the relative parabolic motion. For consistency, we
  have chosen to adopt the same corrections also for the simulations
  run with Chain.}, obtained for a stellar dispersion of 10 km
s$^{-1}$).

After each single-binary encounter we can classify the end-states as
following:\\
\noindent
(A) Fly-by: the binary maintains its components, but it can exit with
a different energy and angular momentum;\\ 
(B) Tidal disruption: the interacting binary is broken by the massive
IMBH. The tidal disruption can end with an ionization (B.1), if the
final system consists of three single bodies, or with an exchange
(B.2), if one of the two components is captured by the single.  The
tidal perturbation occurs at a distance $r_{\rm T}=a_{\rm MSP,i}\left
[M_{\rm {IMBH}}/ (m_{\rm {MSP}}+m_{\rm {WD}})\right]^{1/3}$, where the
gradient exerted by the IMBH on the incoming binary exceeds its
binding energy. For our binary pulsar, $r_{\rm T}\sim 0.1$ AU.

In the case of binary-binary encounters the possible end-states are
analogous (i.e. fly-bies and tidal disruptions), but complicated by
the fact that the interacting binaries are two. In particular, we can
observe the tidal disruption of only one of the two binaries (mostly
the softer [MSP,WD] binary), or of both of them.  After the tidal
disruption of the [MSP,WD] binary:\\
(B.1) The [MSP,WD] can be fully ionized (i.e. both components escape);\\
(B.2) One of the two components remains bound to the [IMBH, star] or
[IMBH,BH] binary, forming a stable/unstable triplet.  Some triplets
show a characteristic configuration of two nested binaries, where two
of the three components are bound in a tight binary, while the other
one orbits around. This type of systems are termed hierarchical
triplets.

A hierarchical triple is stable if it satisfies the relation (Mardling
\& Aarseth 1999) 
\begin{equation}\label{eq:stab}
\frac{R_{\rm p}}{a_{\rm
{in}}}\geq2.8\left[\left(1+q\right)\frac{1+e_{\rm {ou }}}{
\sqrt{1-e^2_{\rm ou }}}\right]^{2/5},
\end{equation}
\noindent
where $R_{\rm p}$ is the pericenter of the outer binary, $a_{\rm in}$
the semi-major axis of the inner binary, $e_{\rm ou }$ the
eccentricity of the outer binary and $q$ the mass ratio between the
external component and the inner binary.  If the triplet is unstable,
the evolution of the system ends with the expulsion of one of the
three components (preferentially, the less bound companion).

In the simulations, the integration is halted when the outgoing
unbound star(s) is (are) at a sufficiently large distance from the
center of mass of the target binary or of the newly formed binary (or
triplet).  This maximum distance has been chosen equal to $50$ times
the semi-major axis of the system left.  If the outgoing star (or
binary) is still at such a distance after at least 2000 time--units,
we stop the integration and we classify the encounter as an unresolved
resonance.

\section{[IMBH,MSP] BINARIES}

\subsection{Cross Sections}

\begin{table*}
\begin{center}
\begin{flushleft}
\begin{tabular}[h!]{|ccccccccccccc|}
  \hline
  $M$ ($\msun$) & $f_{\rm MSP}$ (\%) & $f_{\rm  WD}$ (\%) & 
  $\Sigma_{\rm MSP}$ (AU$^2$) & $\Sigma_{\rm WD}$ (AU$^2$)  \\
  \hline
  [100]  &7.1 &  5.6&223 & 176 \\
  \hline
  [300]  &11.2 &10&350 &315 \\
  \hline
  [100,star]  &3  & 0.63(tr,in)&440 &92 \\
  \hline
  [300,star]   &0.8   &0.15(tr,in)&157 &28\\
  \hline
  $[100,10]_{\rm {h,*}}$ &0.06(tr,in)  & 0.46(tr,in) &3.6 &27\\
  \hline
  $[300,10]_{\rm {h,*}}$  &- &0.16(tr,in) & - &19 \\
  \hline
  $[100,10]_{\rm gw}$  &0.19(tr,ou) &0.04(tr,ou)& 2.4 &0.5 \\
  \hline
  $[300,10]_{\rm gw}$  &0.26(tr,ou)  &0.04(tr,ou)& 36 & 5.5 \\
  \hline
  \hline
  ${\rm [100,star]_{\rm MSP, single}}$  &1.6  & -&126 &- \\
  \hline
  ${\rm [300,star]_{\rm MSP, single}}$   &0.65   &-&66 &-\\
  \hline
\end{tabular}
\end{flushleft}
\end{center}
\label{tab:fraz2}
\caption{Occurrence fractions ($f_{\rm MSP}$ and $f_{\rm WD}$) and
cross sections ($\Sigma{}_{\rm MSP}$ and $\Sigma{}_{\rm WD}$)
calculated from equation (5) of [IMBH,MSP] and [IMBH,WD] binaries for
each initial state of the IMBH, and for PSR-A like MSP binaries.
Bracket (tr,in) denotes the occurrence of stable triplets where the
MSP or the WD binds forming the inner binary. Bracket (tr,ou) denotes
the occurrence of stable triplets where the MSP or WD binds forming
the outer binary.  The last two lines correspond to exchanges of a
single MSP off the [IMBH,star] binary.}
\end{table*}

\begin{table*}
\begin{center}
\begin{flushleft}
\begin{tabular}[h!]{|ccccccccccccc|}
  \hline 
  $M$ ($\msun$) & $w_{\rm X}$& $\Gamma_{\rm MSP}$ ($10^{-11}$yr$^{-1}$) &
$t_{\rm life}$ ($10^8$ yr)\\
  \hline [100] & 0.27&0.3&1.3\\
  \hline [300] &0.27 &0.4  &0.687 \\ 
  \hline [100,star]&0.4 &0.7 &3.6 \\ 
  \hline [300,star] & 0.4&0.3   &2.35\\ 
  \hline \hline ${\rm [100,star]_{MSP, single}}$ &0.2 &0.1& 4.3 \\ 
  \hline ${\rm [300,star]_{MSP, single}}$ &0.2 &0.06 &3.3\\
  \hline
\end{tabular}
\end{flushleft}
\end{center}
\label{tab:fraz3}
\caption{Probability coefficient $w_{\rm X}$ as defined in Section 6,
rates of formation of observable [IMBH,MSP] binaries, and
lifetimes $t_{\rm life,MSP},$ for $\langle v_*\rangle $=10
km${\rm s}^{-1}$, $\langle \rho_*\rangle =7\times 10^5 \,\msun\, {\rm
pc}^{-3}$. The channels of formation are the same as in Table 1.}
\end{table*}

We are interested in deriving the frequency of encounters ending with
the formation of a [IMBH,MSP] binary.  Thus, we computed $f_{\rm
X}\equiv N_{\rm X}/N,$ i.e. the probability factor associated to
channel X, where $N$ is the total number of runs, and $N_{\rm X}$ is
the number of cases in which event $X$ occurs.  The cross section for
channel X can be written as
\begin{equation}
\Sigma_{\rm X}=\pi f_{\rm X}\,b^2_{\rm {max}},
\end{equation}
where $b^2_{\rm {max}}$ is the square of the maximum
impact parameter that includes ``all'' relevant encounters leading to
X (Sigurdsson \& Phinney 1993;
see Section 3 of Appendix A for its operative definition). 
Table 2 summarizes our results.
\begin{itemize}
\item[$\bullet$] In the encounters between the single IMBH and the
[MSP,WD], we find that ionization of the incoming binary leads to the
formation of [IMBH,MSP] systems with an occurrence $\sim{}$10\%.  The
cross section in physical units is about a few hundreds AU$^2$, and
increases with the IMBH mass.
\item[$\bullet$] [IMBH,star]: In the case of binary-binary encounters
with the target binary [IMBH,star], we often observe the exchange
between the star and the heavier MSP, leading to the formation of an
[IMBH,MSP] binary.  The cross section for the formation of the
[IMBH,MSP] binary is slightly larger than for the isolated IMBH in the
case of an IMBH of $100\,\msun,$ whereas the opposite holds for an
IMBH of $100\,\msun.$
\item[$\bullet$] ${\rm [IMBH,star]_{\rm MSP\,\,single}}$: In the
encounter of the [IMBH,star] and the single MSP we again observe the
exchange of the star with the MSP, thus forming an [IMBH,MSP]
system. We note that the frequency is a factor somewhat lower for the
single MSP than in the [MSP,WD] case and this involves smaller cross
sections too.
\item[$\bullet$] [IMBH,BH]: In general, the presence of a massive
companion such as a stellar-mass BH does not favor the formation of an
[IMBH,MSP], since the exchange probability is negligible.  Triple
systems may alternatively form.  In rare cases ($\lsim 0.1\%$) stable
triplets can form with the MSP member of the inner binary
[(IMBH,MSP),BH]. This occurs when the IMBH binary is in its hardening
phase by dynamical encounters.  When the [IMBH,BH] is in the phase of
hardening by emission of gravitational waves, the MSP binds to the
[IMBH,BH] as external companion with an higher probability ($f_{\rm X}
\sim 0.2-0.3\%$) than in the hardening by scattering regime.
\end{itemize}

\subsection{[IMBH,MSP] binary parameters}

\begin{figure}
\begin{center}
\centerline{\psfig{figure=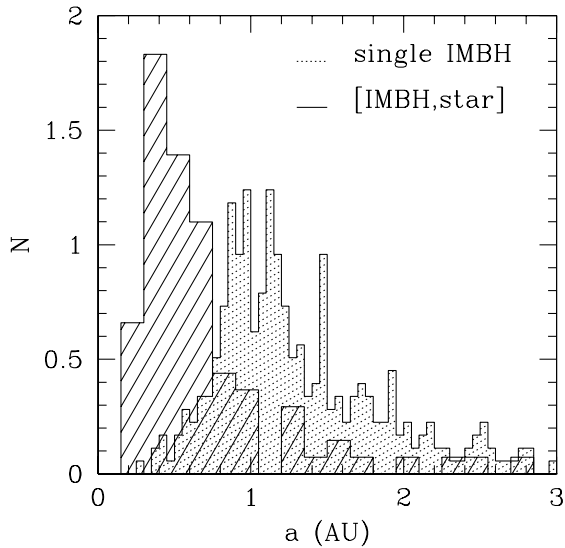,height=8.cm}}
\caption{Distribution of the semi-major axes of the [IMBH,MSP]
binaries, normalized to the corresponding fraction of events, for
PSR-A like initial MSP binaries. The IMBH has a mass of $100\,\msun.$ 
Shaded histogram with dotted lines refers to [IMBH,MSP] systems formed
after tidal disruption off the single IMBH. Shaded histogram with
solid lines refers to the [IMBH,MSP] binaries that form after the
exchange of the initial star in the [IMBH,star] binary.}
\label{fig:fig1}
\end{center}
\end{figure}

\begin{figure}
\begin{center}
\centerline{\psfig{figure=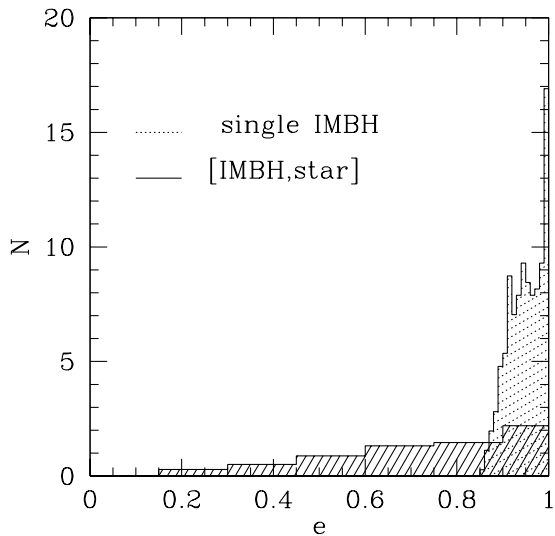,height=8.cm}}
\caption{Distribution of eccentricities of [IMBH,MSP] binaries,
normalized to the corresponding fraction of events. Shaded histograms
refer to the same cases as in Fig.~\ref{fig:fig1}.}
\label{fig:fig2}
\end{center}
\end{figure}

In this section we explore the properties of the [IMBH,MSP] systems
that have formed dynamically. Fig. \ref{fig:fig1} shows the
distribution of semi-major axes resulting from encounters with the
$100\, \msun$ IMBH.  In the case of tidal disruption of the [MSP,WD]
off the single IMBH, we find that the distribution peaks at $\sim{}$1
AU.  This value agrees with the analytical estimate (Pfahl 2005)
obtained in the impulse approximation, i.e. considering that the
incoming [MSP,WD] binary is approaching the IMBH along a parabolic
orbit, and that is disrupted instantaneously at the tidal radius
$r_{\rm T}.$ According to this analytical model (Pfahl 2005), the most
likely end-state has a binding energy per unit mass
\begin{equation}
E\sim  -{m_{\rm WD}\over m_{\rm MSP}+m_{\rm WD}}V_{\rm T}V_{\rm rel}
\end{equation}
where $V_{\rm T}\sim (GM_{\rm IMBH}/r_{\rm T})^{1/2}$ and $V_{\rm
rel}$ is the relative velocity of the [MSP,WD] binary before the
encounter.  The corresponding semi-major axis of the newly formed
[IMBH,MSP] binary is
\begin{equation}\label{eqn:apfahl}
a_{\rm {MSP,f}}\sim { a_{\rm {MSP,i}}\over 2\sqrt{2}}{M_{\rm
IMBH}\over m_{\rm WD}} \left({m_{\rm MSP}+m_{\rm WD}}\over M_{\rm
IMBH}\right )^{1/3}
\end{equation}
which perfectly agrees with the results of our simulations 
($a_{\rm {MSP,f}}\sim 1$ AU for a $100\, \msun$ IMBH).

Fig. \ref{fig:fig1} also shows the distribution of the semi--major
axes of the [IMBH,MSP] formed during the [MSP,WD] interaction off the
[IMBH, star] binary, following the disruption of the [MSP,WD] at
$\sim{}r_{\rm T}$ and the subsequent exchange of the MSP off the star.
The MSP is captured on a close orbit, and, from simple energy
arguments, the most likely end--state is expected to have a specific
energy
\begin{equation}
E\sim  -{m_{\rm WD}\over m_{\rm MSP}+m_{\rm WD}}V_{\rm T}V_{\rm rel} - 
\frac{m_*}{ a_*} {GM_{\rm IMBH}\over 2m_{\rm MSP}}. 
\end{equation}
Indeed, during the triple encounter between the MSP, the star and the
IMBH (after the expulsion of the WD), an energy (at least) equal to
the binding energy of the star before ejection needs to be extracted,
in order to unbind the star.  The characteristic semi-major axis of
the newly formed [IMBH,MSP] will thus be
\begin{equation}
a^*_{\rm MSP,f}\sim {a_{\rm {MSP,f}}\over 1 + 
({m_*/m_{\rm MSP}})\,\,{a_{\rm {MSP,f}}/ a_*}}.
\end{equation}
If we consider mean values for the initial $m_*/a_*$ selecting all the
systems that end with an [IMBH,MSP] binary, we find $m_*/a_*\sim 1.68
\,\msun/\,\rm {AU}.$ This corresponds to a semi-analytical estimate
$a^*_{\rm MSP,f}\sim$ 0.45 AU, in good agreement with the peak of the
corresponding semi-major axis distribution derived from our
simulations (Fig.~\ref{fig:fig1}).

Fig. \ref{fig:fig2} shows the distribution of the eccentricities for
the same binaries.  For the case of tidal capture the eccentricities
at which the MSP binds to the IMBH are above 0.9; for the formation
channel through exchange the spread of the final eccentricity
distribution is much larger, according to a thermal distribution. This
can eventually be the effect of repeated interactions between the MSP
and the initial companion of the IMBH during the transient state of
unstable triplet.  The distribution of the semi-major axis and
eccentricity of [IMBH,MSP]$_{\rm MSP\,\,single}$ systems formed by the
exchange off the single MSP are similar to the ones formed in the
interaction of the [MSP,WD] off the [IMBH,star].

Finally we note that in the case of a $300\,\msun$ IMBH, the
distributions are similar and only slightly skewed to larger values of
the semi-major axes, as should be expected for a more massive BH (see
equation \ref{eqn:apfahl}).

\subsection{Hierarchical triplets}

As previously noted, the only way a MSP can be retained in the
presence of an [IMBH,BH] binary is through the formation of
hierarchical stable triple systems. Two possibilities exist: either
the formation of a [(IMBH,MSP),BH] where the MSP is closely bound to
the IMBH, or the formation of a [(IMBH,BH),MSP] with the MSP as
external object.

\begin{figure}
\begin{center}
\centerline{\psfig{figure=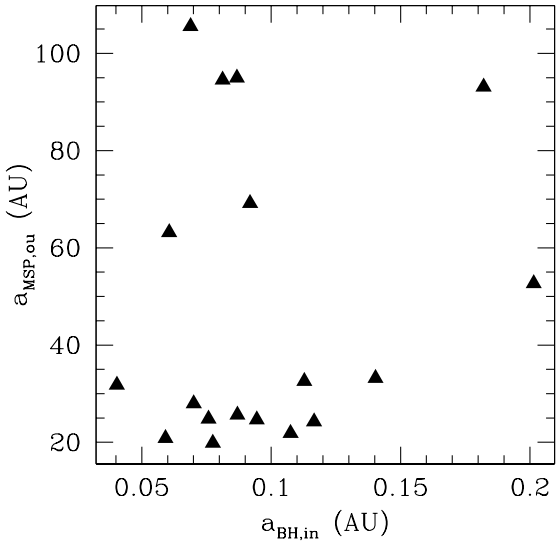,height=8.cm}}
\caption{MSP semi-major axis $a_{\rm MSP,ou }$ of the outer binary
versus semi-major axis $a_{\rm BH,in }$ of the inner binary (IMBH,BH)
of stable hierarchical triple systems. The plot refers to an initial
[IMBH,BH] binary of $100\,\msun$ and $10\,\msun,$ and a initial
PSR-A-like MSP binary.}
\label{fig:fig3}
\end{center}
\end{figure}

\begin{figure}
\begin{center}
\centerline{\psfig{figure=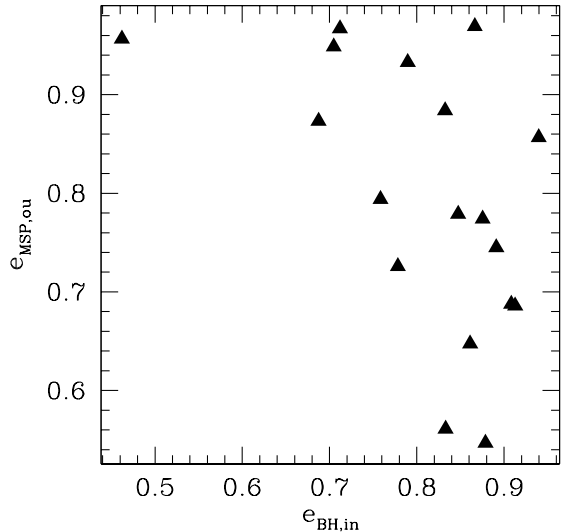,height=8.cm}}
\caption{MSP eccentricity $e_{\rm MSP,ou }$ of the outer binary versus
eccentricity $e_{\rm BH,in}$ of the inner binary (IMBH,BH) of stable
hierarchical triple systems: the initial parameters of the involved
binaries are the same as in Fig. \ref{fig:fig3}.}
\label{fig:fig4}
\end{center}
\end{figure}

Triple systems of the first type are rare, because the MSP tends to
bind preferentially on orbits where its motion is gravitationally
perturbed by the stellar-mass BH causing the MSP to be finally
ejected.  Only triplets of the second type are seen to form with a non
negligible probability ($\sim 0.2 $\%): the MSP binds on very wide
(20-100 \, AU), eccentric orbits ($>0.6$), as shown in Figs.
\ref{fig:fig3} and \ref{fig:fig4}.  The triplets in consideration are
extremely hierarchical (i.e., $R_{\rm MSP,ou }\gg{}a_{\rm BH,in}$), in
order to fulfill the stability condition.

Hierarchical triplets of this type are likely to survive inside the GC
and to turn into a [IMBH,MSP].  Indeed, once the triplet has formed,
the MSP shrinks its orbit with time due to dynamical encounters off
cluster stars while the inner binary hardens due to gravitational wave
emission.  Since the hardening time of the inner binary is usually
shorter than that of the outer binary, these triplets are transient
states ending with the formation of a new [IMBH,MSP] binary following
BH coalescence.

\section {[IMBH,WD] binaries}

\begin{figure}
\begin{center}
\includegraphics[width=7cm]{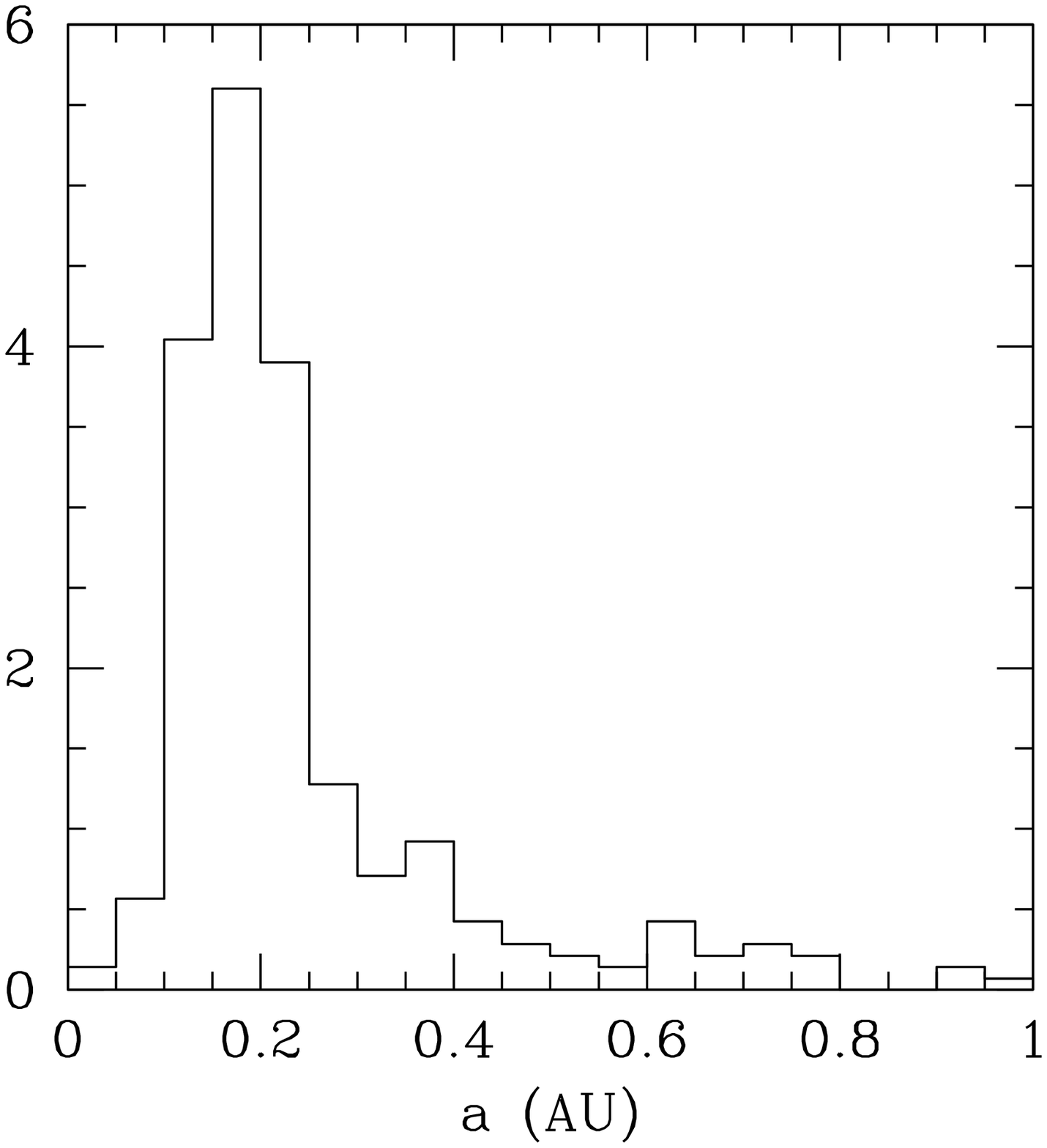} 
\includegraphics[width=7cm]{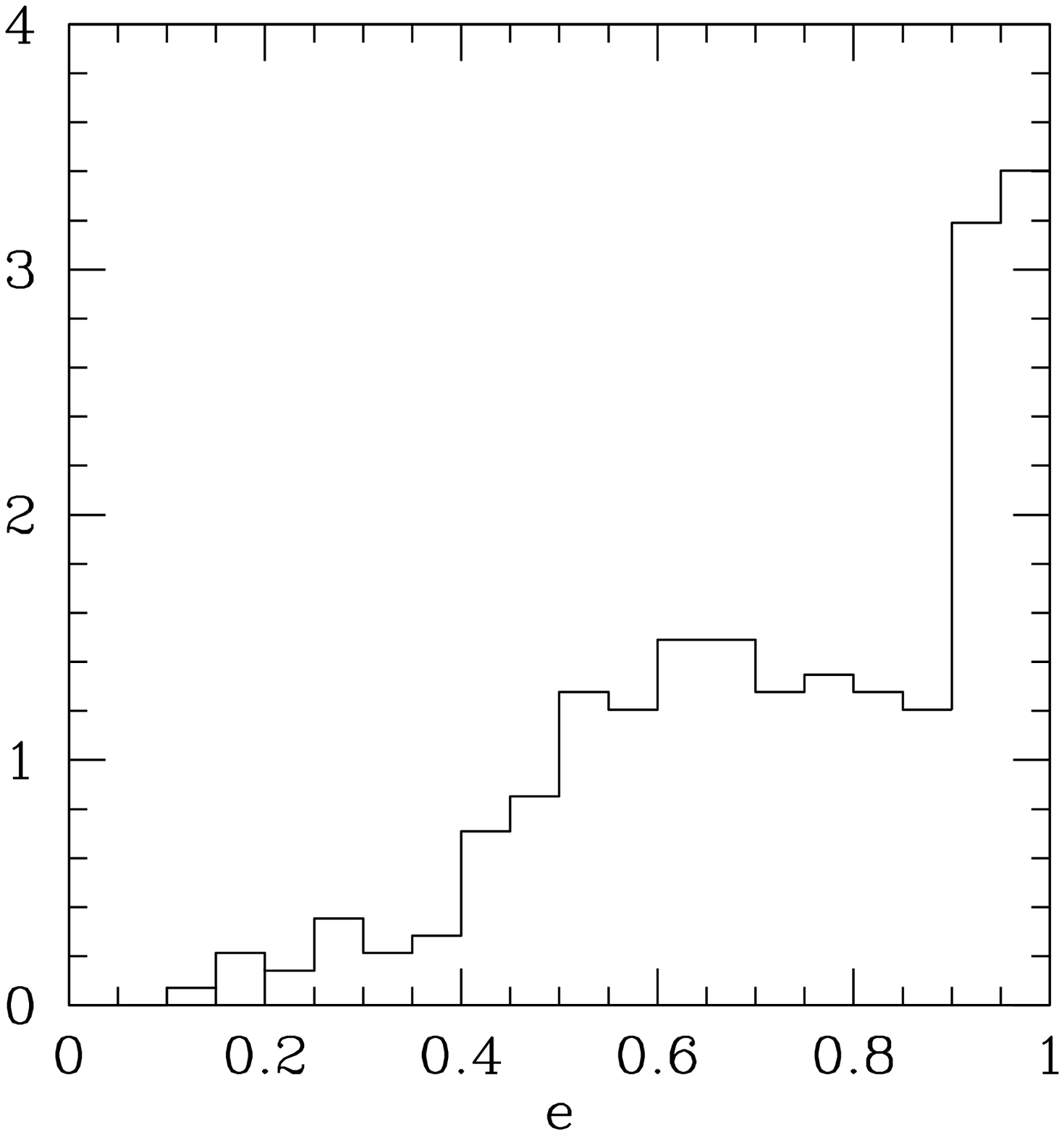}
\caption{Distributions of semi-major axis and eccentricities of the
[IMBH,WD] binaries, normalized to the corresponding fraction of
events, for the single IMBH of $100\,\msun,$ and a PSR-A like initial
MSP binary.}
\label{fig:fig5}
\end{center}
\end{figure}

For the sake of completeness, the results on the formation of
[IMBH,WD] binaries are also summarized in Table 2.  In the case of the
capture of the WD by the single IMBH, we note that the occurrence
fraction of [IMBH,WD] is only slightly lower than that of [IMBH,MSP]
while it decreases of a factor $\sim$ 5 for the [IMBH,star] cases, as
shown in Table 2.  If the IMBH has a companion star, the WD
preferentially binds in triplet configurations. In fact the WD can be
retained around the IMBH only if it forms a hierarchical triplet
[(IMBH,WD),star]. This is due to the smaller mass of the WD relative
to the star that makes exchanges very unlikely.  The same is true for
the [IMBH,BH] cases: stable triplets form with the WD in the inner
binary, i.e [(IMBH,WD),BH], when the IMBH binary is hardening by
scattering stars. On the contrary, the fraction of stable triplets
significantly drops during the gravitational wave driven phase ($\sim
0.04 \%$). This is due to the fact that the WD preferentially binds to
the IMBH on a orbit strongly perturbed by the stellar mass BH.  The
cross sections computed using equation (5) are reported in Table 2 and
their values reflect their dependence upon $f_{\rm X}.$

Fig. \ref{fig:fig5} shows the distributions of the semi-major axis and
eccentricity for the WD case, considering only the interaction with
the single IMBH.  Because of its lighter mass with respect to the MSP,
the WD binds around the single IMBH on tighter orbits and the peak is
around 0.17 AU, in agreement with Pfahl's analysis (2005)\footnote{If
the WD is captured instead of the MSP, equation (\ref{eqn:apfahl}) is
modified to take into account for the different mass of the expelled
star, thus giving $a_{\rm {WD,f}}\sim { a_{\rm {MSP,i}}\over
2\sqrt{2}}{M_{\rm IMBH}\over m_{\rm MSP}} \left({m_{\rm MSP}+m_{\rm
WD}}\over M_{\rm IMBH}\right )^{1/3}=0.14 M^{2/3}_{100}$ AU.}.

The channel that we have outlined for the formation of a [IMBH,WD]
binary is probably not the dominant one, because of the higher number
of [WD,star] with respect to [MSP,WD] binaries. For this reason we
have chosen not to discuss the formation rate of [IMBH,WD] binaries in
more details.

\section{[IMBH,MSP] in globular clusters}

So far, we have considered only binary MSPs which mimic the properties
of PSR-A in NGC~6752. Compared to PSR--A however, binary MSPs in GCs
display a wider distribution of properties in their orbits and masses
(Camilo \& Rasio 2005).  Since the cross section for the formation of
[IMBH,MSP] systems as well as their ending states depend on the
initial semi--major axes and total mass of the impinging [MSP,WD]
binaries, in this section we have simulated a set of interactions
varying the properties of the binary MSP.

Binary MSPs in GCs show a double peaked distribution of their
semi--major axes in the interval $[0.0024 \, \rm {AU},0.035$ AU],
while a number of ``outliers'' spread over larger orbital separations
(see Fig. 3 in Camilo \& Rasio 2005). Outliers count for the 25\% of
the entire population.  We have fitted the observed distribution with
(i) an asymmetric Landau profile, peaked at 0.005 AU, in the range
[0.0024 AU, 0.02 AU] (defining class I [short period binary MSPs]),
plus (ii) a Gaussian profile, centered around 0.026 AU, in the range
[0.02 AU, 0.035 AU] (defining class II [long period binary MSPs]).
According to Camilo \& Rasio (2005), we have assigned a companion WD
mass of $0.03 \,\msun$ for class I, and of $m_{\rm WD}= 0.2\, \msun$
for class II.  For the binary MSPs referred to as outliers, we have
taken $a_{\rm MSP,i}=0.21$ AU and $m_{\rm WD}=0.34\,\msun,$
corresponding to their mean properties.

\subsection{Cross sections}

\begin{table*}
\begin{center}
\begin{flushleft}
\begin{tabular}[h!]{|ccccccccccccc|}
  \hline 
  $M$ ($\msun$)&N&$f_{\rm MSP} (\%)$ 
  &$\Sigma_{\rm MSP} ({\rm AU^2})$ &$ w_{\rm X}$ 
  & $\Gamma_{\rm X}$ ($10^{-11}$yr)&$t_{\rm life}$ ($10^8$ yr)\\
  \hline $[100]_{\rm I+II} $&10000 &10.7&260 &0.2&0.2 &0.6\\
  \hline $[100]_{\rm outlier}$&10000  &10.8&3900 &0.07&1.2 &2.2\\ 
  \hline $[100,{\rm star}]_{\rm I+II}$ & 5000 &1.8  & 232&0.3 &0.3 & 4.3\\ 
  \hline $[100,{\rm star}]_{\rm outlier}$ & 5000  &5.2  &680 &0.1&0.3&5.5\\
  \hline
\end{tabular}
\end{flushleft}
\end{center}
\label{tab:fraz3}
\caption{Outcomes from the encounters of different kinds of binary
  MSPs in GCs with a single or a binary IMBH of $100\,\msun$. Columns:
  number N of runs for each set of simulations
  , occurrence fraction ($f_{\rm MSP}$ normalized
  to N), cross section $\Sigma_{\rm MSP}$ (as defined in Section 3.1),
  probability coefficient $w_{\rm X}$ as defined in Section 6,
  characteristic formation rates $\Gamma_{\rm X}$, and lifetimes $t_{\rm
    life}$ (estimated as in Section $5.3$). These times are computed
  considering $\langle v_*\rangle$=10 km${\rm s}^{-1}$, $\langle
  \rho_*\rangle = 7\times 10^5\, \msun \,{\rm pc}^{-3}$, and a core
  radius of 0.75 pc.  First (last) two rows refer to encounters with
  class I+II binaries and to outliers scattering off a single (binary)
  IMBH, respectively.}
\end{table*}

Table 4 collects the results obtained considering as target an IMBH of
$100\,\msun$.  We find, in the case of the single IMBH, that the cross
section is larger for the outliers compared to class I+II, due to
their initially wider separation. 
For the [IMBH,MSP] binaries formed following the exchange of the
initial stellar companion we obtain similar results, but the
differences in cross section between outliers and class I and II is
less pronounced.

\subsection{Orbital parameters} 

\begin{figure}
\begin{center}
\includegraphics[width=7cm]{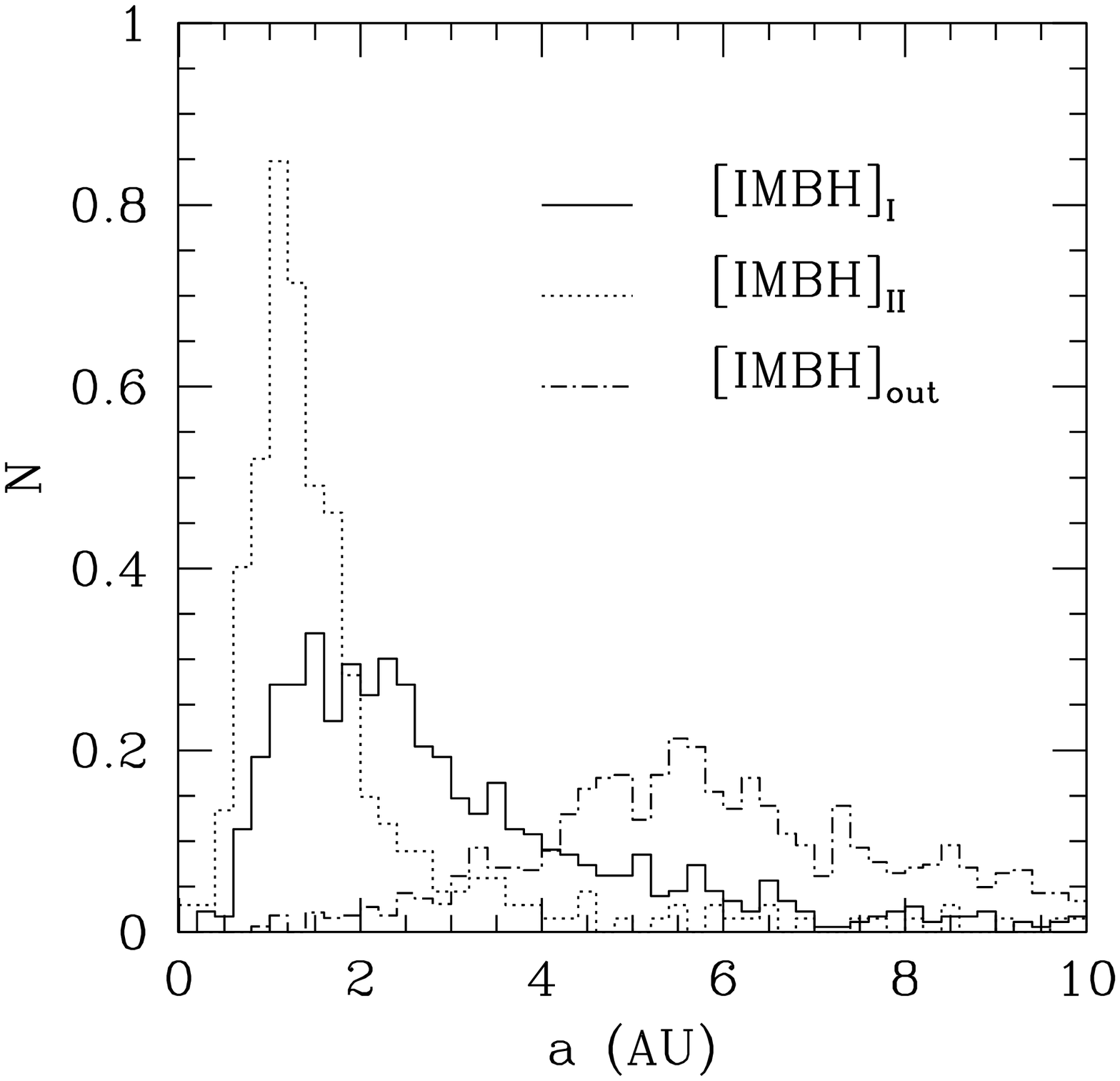} 
\includegraphics[width=7cm]{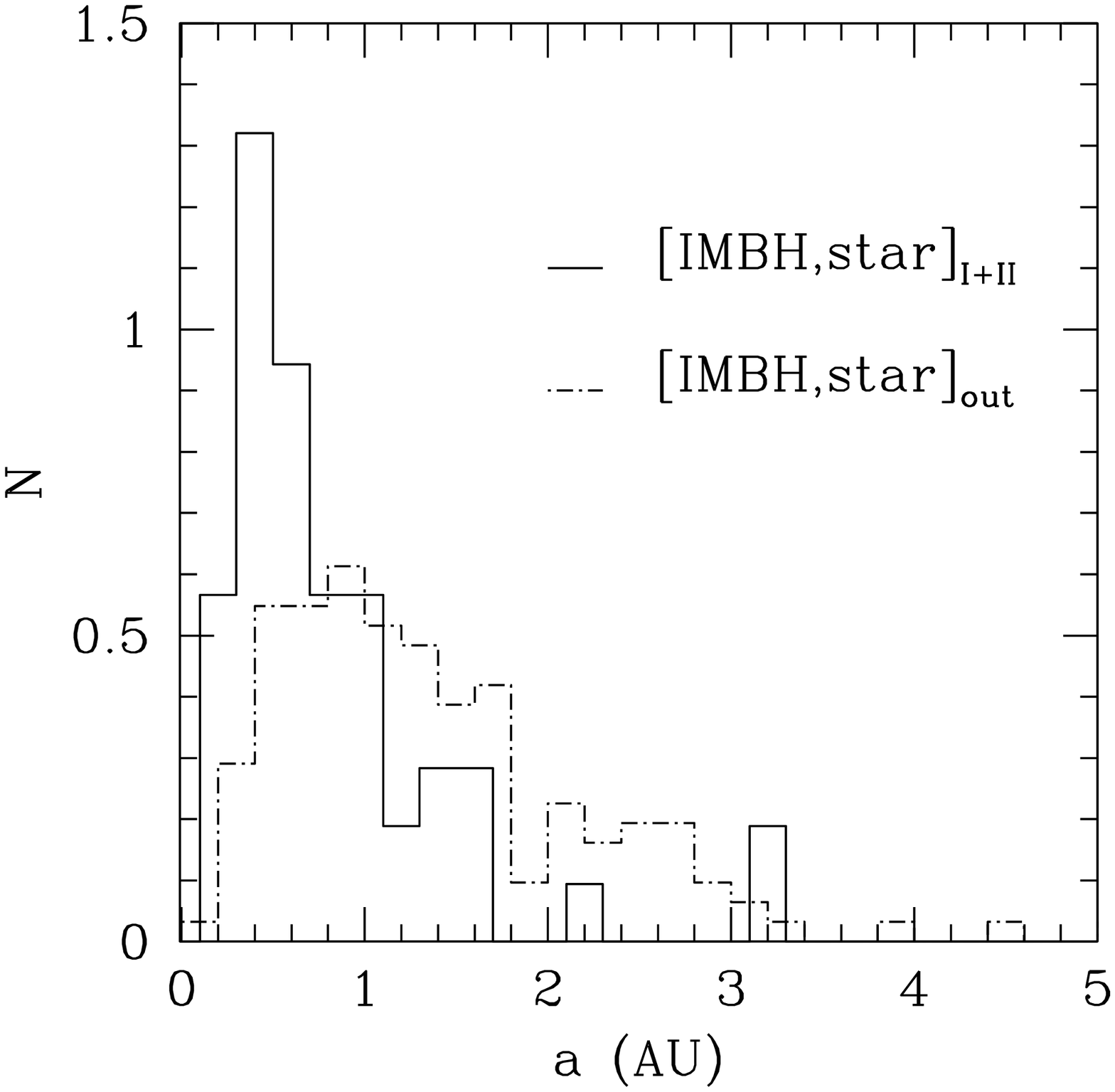}
\caption{ Distribution of the semi-major axes of [IMBH,MSP] binaries,
  normalized to the corresponding fraction of events.  The IMBH has a
  mass of $100\,\msun.$ Left panel refers to encounters off the single
  IMBH; solid, dotted and dot-dashed, lines refer to scattering with
  class I, class II and outliers, respectively.  Right panel refers to
  encounters off the [IMBH,star] binary: solid, and dashed lines refer
  to class I+II, and outliers, respectively.}
\label{fig:fig6}
\end{center}
\end{figure}

\begin{figure}
\begin{center}
\includegraphics[width=7cm]{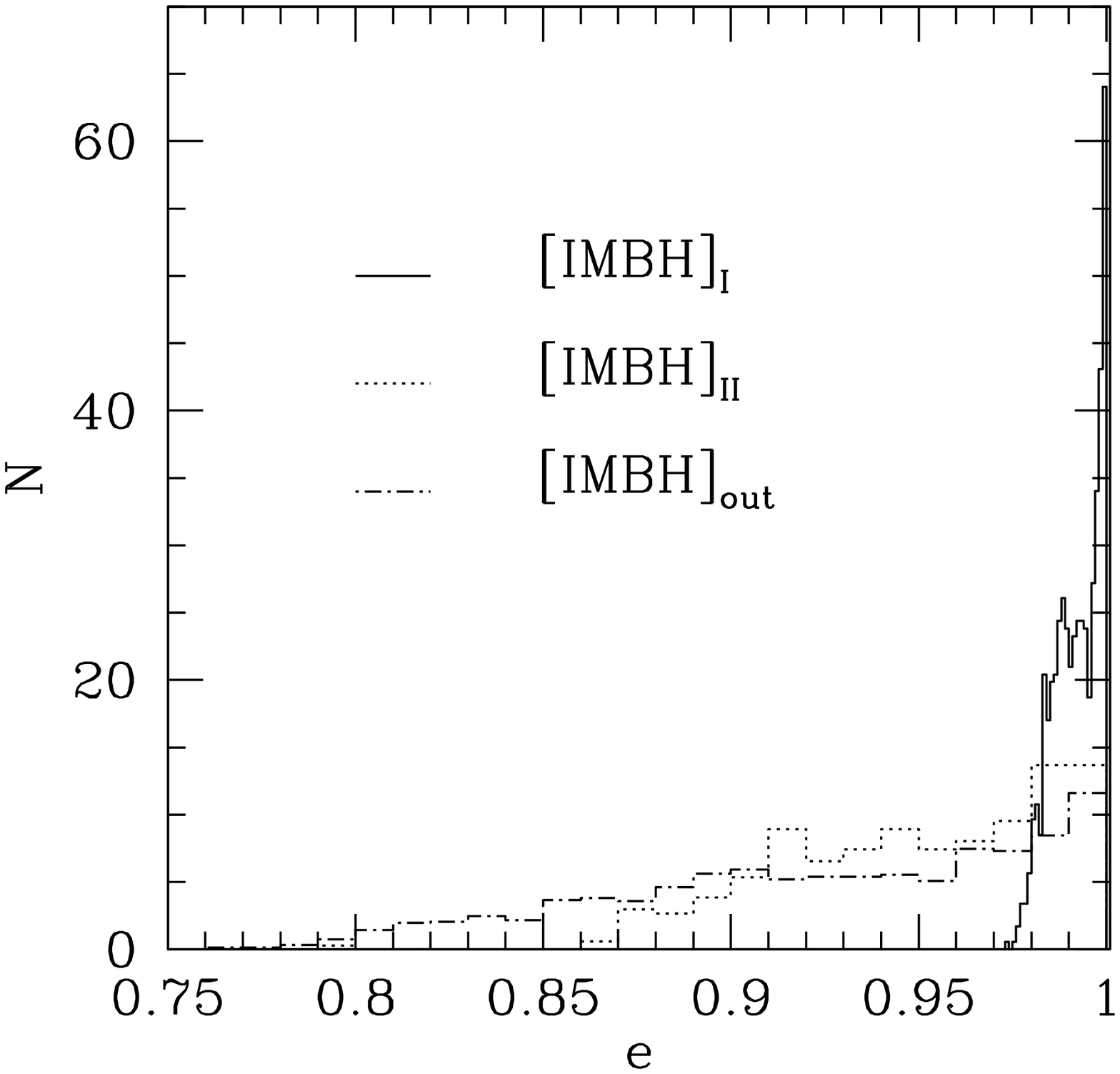} 
\includegraphics[width=7cm]{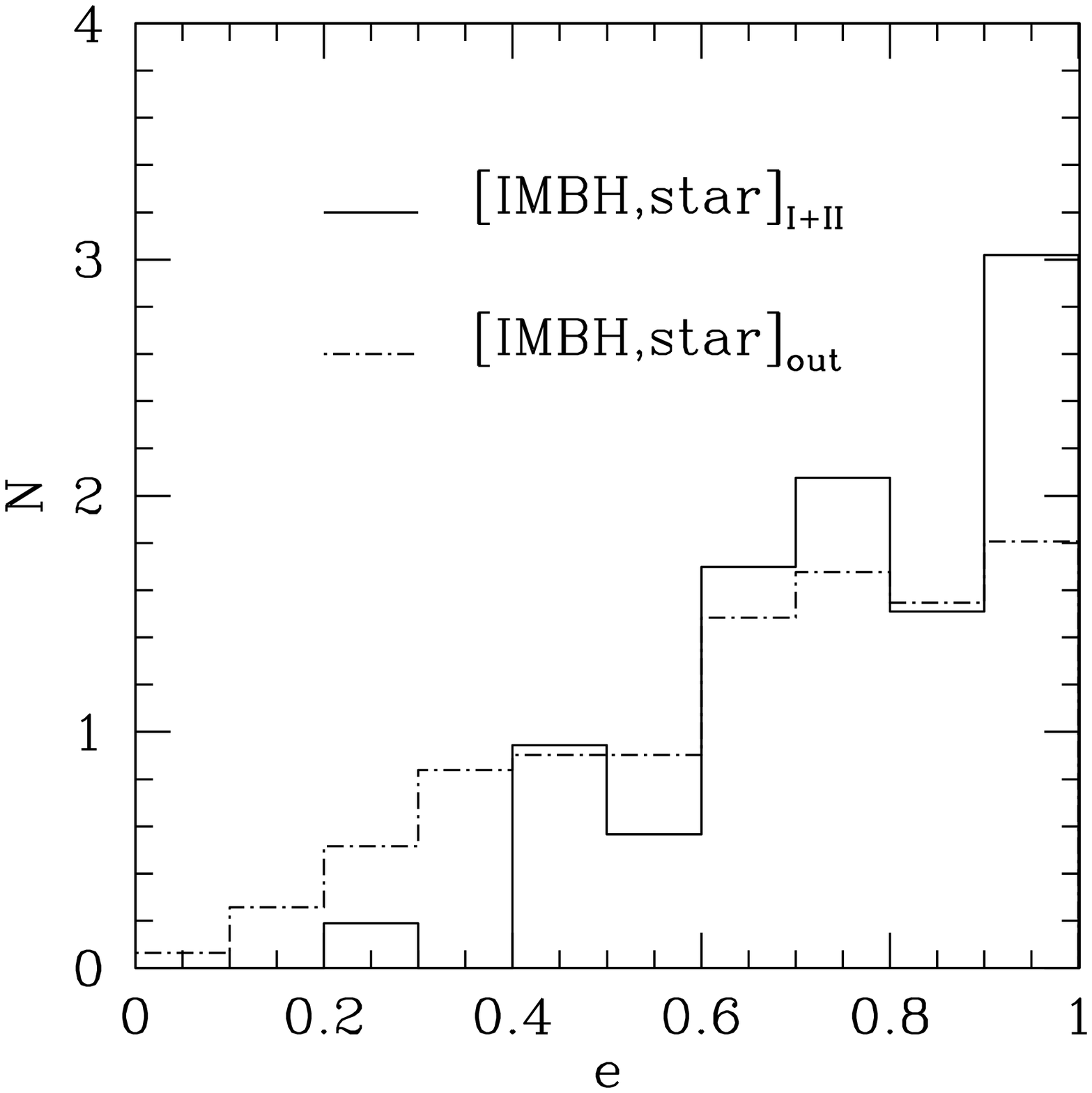}
\caption{ Distribution of the eccentricities of [IMBH,MSP] binaries,
  normalized to the corresponding fraction of events. The IMBH has a
  mass of $100\,\msun.$ Left panel refers to encounters off the single
  IMBH; solid, dotted and dot-dashed, lines refer to scattering with
  class I, class II and outliers, respectively.  Right panel refers to
  encounters off the [IMBH,star] binary: solid, and dashed lines refer
  to class I+II, and outliers, respectively.}
\label{fig:fig7}
\end{center}
\end{figure}

Fig. \ref{fig:fig6} (left panel) shows the distributions of the
semi-major axes of the [IMBH,MSP] binaries formed after the
interactions off a single IMBH. It appears that different populations
of [MSP,WD] binaries lead to the formation of [IMBH,MSP] systems with
different orbital characteristics.  The peak of the semi-major axis
distribution for each class can be inferred from equation (7): 1.7 AU
for the short period, class I binaries, 1.1 AU for the long period,
class II binaries, and 5.6 AU for the outliers.  A clear trend is also
visible for the eccentricities (Fig. \ref{fig:fig7} left panel): the
lighter the WD is, the more eccentric (and with a narrower spread) is
the orbit of the [IMBH,MSP] binary.  This correlation is due to
angular momentum conservation:

\begin{center}
\begin{equation}
m_{\rm WD}\sqrt{\frac{{\rm G}a_{\rm MSP,i}}{m_{\rm MSP}+m_{\rm WD}}}=  
M_{\rm IMBH}\sqrt{\frac{{\rm G}a_{\rm MSP,f}(1-e^2_{\rm f})}
{m_{\rm MSP}+M_{\rm IMBH}}}. 
\end{equation}
\end{center}  
Using equation (7) this implies 
\begin{center}
\begin{equation}
1-e^2_{\rm f} \propto m^3_{\rm WD}(m_{\rm MSP}+m_{\rm WD})^{-4/3 }. 
\end{equation}
\end{center}

Fig. \ref{fig:fig6} (right panel) shows the semi-major axes of the
[IMBH,MSP] systems formed after the interaction with the [IMBH,star]
systems.  The distributions are skewed to smaller separations,
compared to the case of a single IMBH, due to the fact that the MSP
has ejected the star (see Section 3.2).  The smaller cross section for
the [IMBH,star] case compared to the single IMBH, for the family of
the outliers (see Table 4), is due to the occurrence of unstable
triplets where the MSP, that binds onto wider orbit (see equation
(7)), is preferentially expelled.  Fig. \ref{fig:fig7} (right panel)
shows the eccentricity distribution, relative to encounters off the
[IMBH,star] binaries, which it turns out similar to that of Fig.
\ref{fig:fig2}.

\subsection {Lifetimes}

\begin{figure}
\begin{center}
\includegraphics[width=7cm]{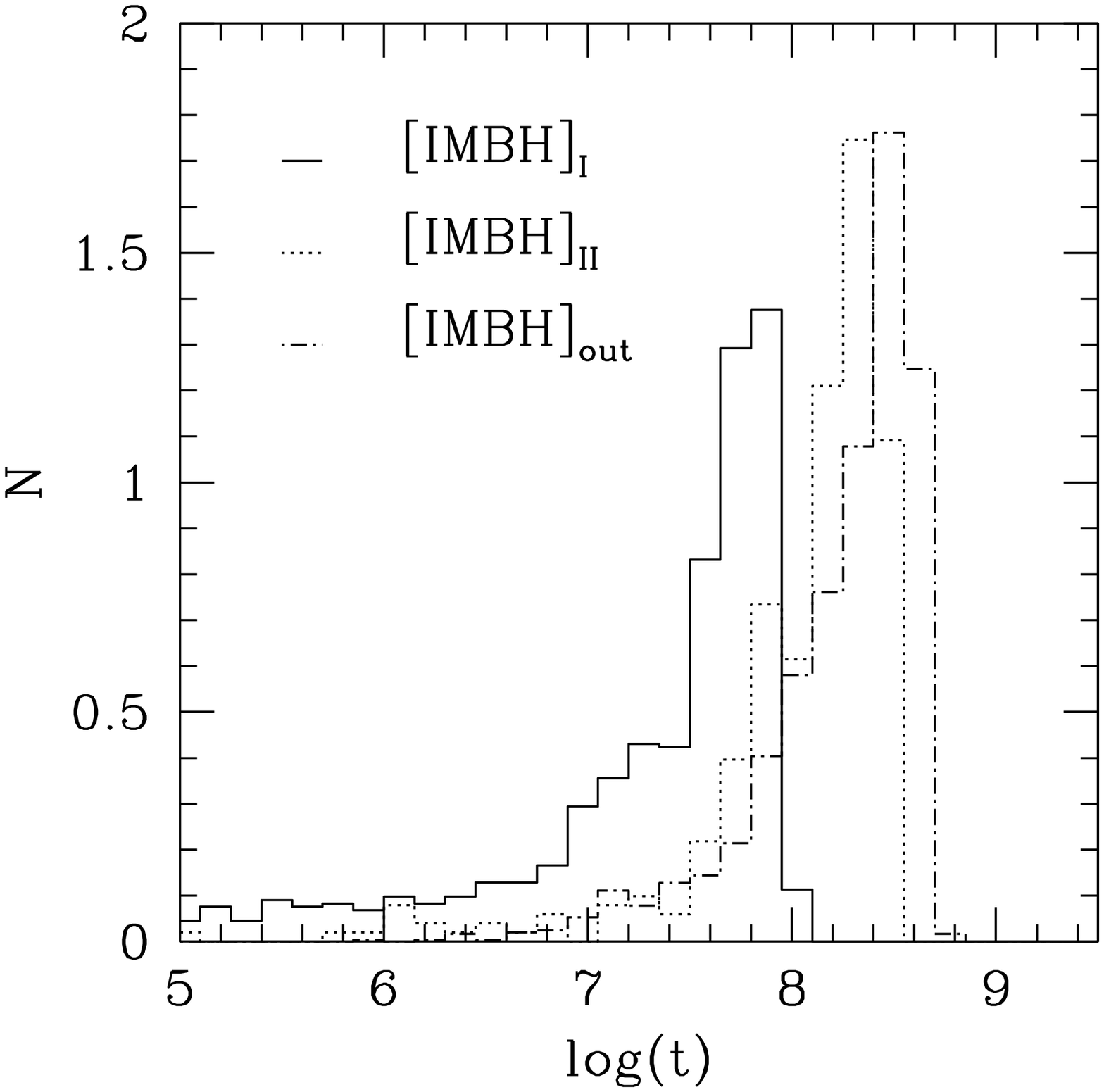} 
\includegraphics[width=7cm]{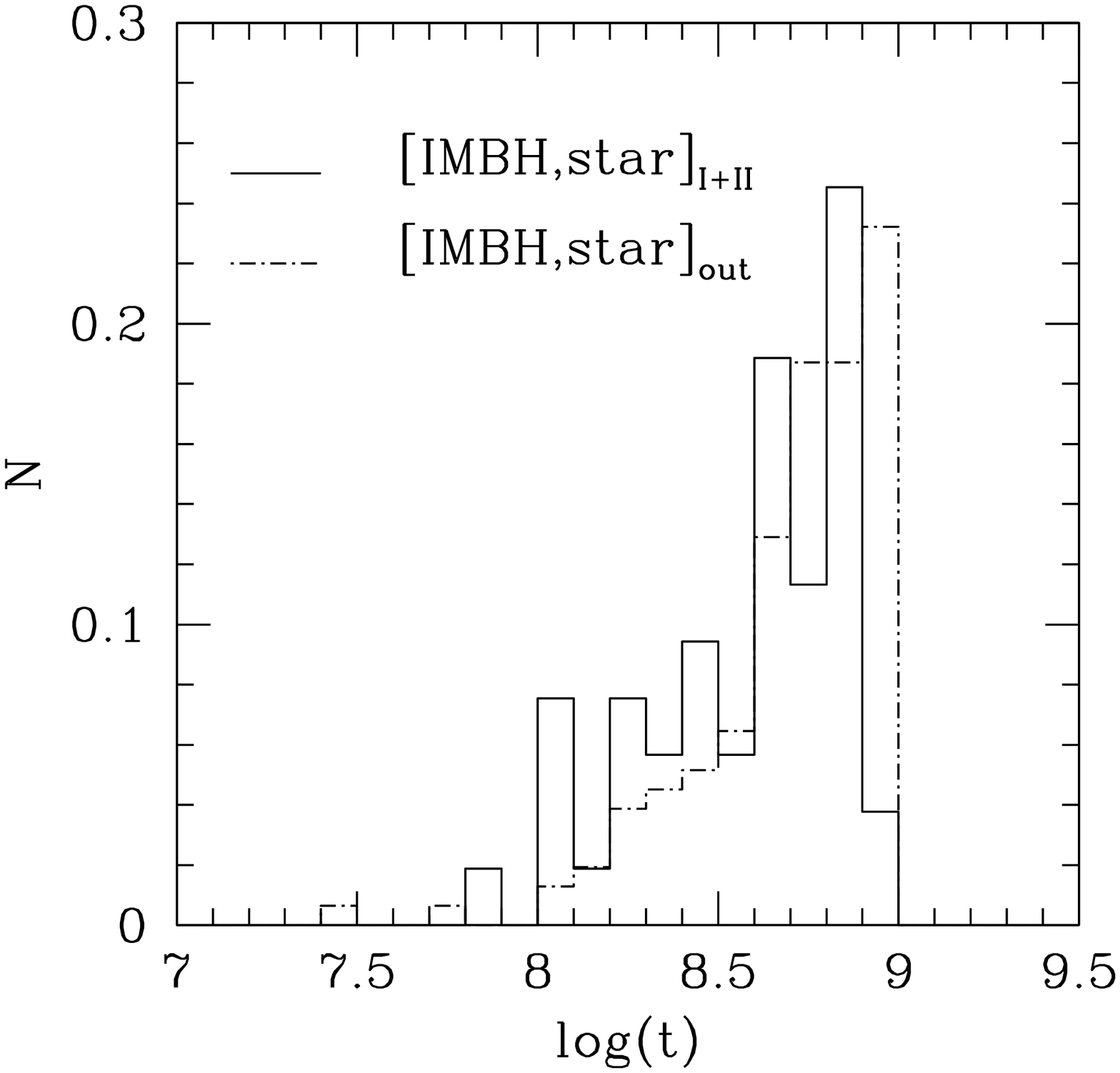}
\caption{Distribution of the lifetimes.  Lines and labels are defined
as in Fig. \ref{fig:fig6}}
\label{fig:fig8}
\end{center}
\end{figure}

The simulations provide the semi-major axes and eccentricities of the
[IMBH,MSP] systems formed. So, using equations (\ref{eq:aharde}),
(\ref{eqn:dagw}) and (\ref{eqn:degw}) of Section 1 of Appendix A, we
can calculate their subsequent orbital evolution, controlled either by
hardening off cluster stars or by gravitational wave back--reaction.
The lifetime is defined as the sum of the time necessary for the
individual binary to harden by stars until the separation $a_{\rm gw}$
(equation (3)) is attained, plus the time for gravitational wave
in-spiral at $a_{\rm gw}$ , i.e., $t_{\rm life}=t_{\rm h}+t_{\rm gw}$.
The mean values of the binary lifetimes are reported in Table 3 for
PSR-A-like initial MSP binaries, and in Table 4 for the complete
population. Note that $t_{\rm life}$ is computed assuming that the
eccentricity $e_{\rm MSP}$ does not vary during the hardening phase
against stars.  A further increase in $e_{\rm MSP}$ can bring the
binary into the gravitational waves regime faster, while a reduction
can make the binary more long-lived. The [MSP,IMBH] binaries formed
are already very eccentric. If dynamical interaction tends to bring
the eccentricity distribution closer to the thermal one, we then can
argue that our estimated lifetimes represent lower limits.

Fig. \ref{fig:fig8} shows the characteristic lifetimes of the
[IMBH,MSP] binaries described in Section 5.2.  Left panel refers to
encounters off the single $100\,\msun$ IMBH. We note that the different
families of [MSP,WD] binaries lead to [IMBH,MSP] systems with
different lifetimes: in particular for class I, $t_{\rm life}\approx
6\times 10^7$ yr due to the extremely high eccentricities at which
the new systems form.  By contrast, class II and the outliers have
higher $t_{\rm life}\gsim 10^8$ yr.  Right panel of Fig.
\ref{fig:fig8} refers to encounters of MSP binaries off the
[IMBH,star] system.  In this case the distributions seem not to depend
strongly on the incoming binaries: outliers as well as class I and II
show very similar lifetime distributions with characteristic values
around $4\times 10^8$ yr.

\section{Detectability of [IMBH,MSP] binaries in globular clusters}

In Section 3 and 5 we investigated the formation of binaries hosting
an IMBH and a MSP, via single--binary and binary--binary interactions.
Here, we compute their formation rates and estimate 
the number of expected systems in the Milky Way GCs. 

The rate of formation for channel X reads  
\begin{center}
\begin{equation}
\Gamma_{\rm X}\sim n_{\rm MSP} w_{\rm X}
\langle v_{ \infty}\rangle \Sigma_{\rm X} 
\label{gamma}
\end{equation}
\end{center}

\noindent
where $n_{\rm MSP}$ is the number density of MSPs in the cluster core
of radius $r_{\rm c}$, $\Sigma_{\rm X}$ the cross section defined in
equation (5) and $w_{\rm X}$ the probability coefficient (estimated
below), associated to channel X.

The structural parameters of GCs span a large interval of values. In
order to estimate $\Gamma_{\rm X},$ we considered only the 23 GCs that
are known to host at least one MSP.  For each GC in this selected
sample, we computed the MSP number density as $n_{\rm MSP}\sim N_{\rm
  MSP}/4 r_{\rm c}^3$ where $N_{\rm MSP}$ is half of the number of
currently observed MSPs in every GC in order to take into account the
fact that not all MSPs are hosted inside the GC's core. The mean value
of $n_{\rm MSP}$ obtained considering the sample of galactic GCs is
$\approx 2 \times10^{-14}$ AU$^{-3}$.

For the calculation of $w_{\rm X},$ we adopted a ratio of 2 for the
relative number of single and binary MSPs, in accordance with the
ratio observed (Camilo \& Rasio 2005). The outliers account for 25\%
of the binary MSPs, and class I and II for 50\% and 25\%,
respectively.  Following Blecha et al. (2006), we also assume that the
IMBH lives as single object for $\sim 40\%$ of its lifetime, whereas
for the remaining $\sim 60\%$ it is bound with a cluster star. The
values of $w_{\rm X}$ are computed according to these simple recipes
and are collected in Tables 3 and 4 together with the estimates mean
rates $\Gamma_{\rm X}$. We note that the main contribution comes from
binary MSPs belonging to the family of the outliers, scattering off
the single IMBH.
 
As previously discussed in Section 5.3 and shown in Fig. 8, the
[IMBH,MSP] binaries have characteristic lifetimes shorter than their
typical formation timescales.  Consequently, the expected number of
[IMBH,MSP] binaries that formed and reside in a GC is roughly given by
\begin{center}
\begin{equation}
N_{\rm X} \sim t_{\rm life,X} \Gamma_{\rm X}.
\label{nx}
\end{equation}
\end{center}
We thus estimated the total number $N^{\rm exp}_{\rm tot}$ of expected
[IMBH,MSP] systems (i.e. those [IMBH,MSP] in which the radio beams of
the MSP sweep the direction to the Earth), summing over all channels
and over the sample of GCs hosting at least one MSP.  We find $N^{\rm
  exp}_{\rm tot}\sim 0.1,$ if a $\sim 100\,\msun$ IMBH is hosted in
{\it all} the GCs which are currently known to include a MSP. Thus,
the detection of an [IMBH,MSP] binary has at present a low probability
of occurrence \footnote{ No strong bias against the detection of an
  [IMBH,MSP] binary is caused by its the orbital motion. In fact,
  Patruno et al. (2005) showed that the discovery of a bright MSPs
  orbiting an IMBHs at mean separations of a few AU is not hampered by
  the Doppler modulation of the radio pulses.}.

The derived value of $N^{\rm exp}_{\rm tot}$ is a firm lower limit
since $n_{\rm MSP}$ represents a lower limit to the MSP density in a
GC core, given that we considered only the already detected MSPs.  The
ongoing deep surveys running at GBT (Ransom et al. 2005), GMRT (Freire
et al. 2004) and Parkes (Possenti et al. 2003) are rapidly increasing
the known population of MSPs in GCs, suggesting that additional
clusters may contain a rich population of MSPs. The likelihood of
unveiling a binary [IMBH,MSP] will become significantly higher when
new more powerful radio telescopes will become available. In
particular the planned SKA (Cordes et al. 2004) is expected to improve
of 1-2 orders of magnitude the sensitivity limits of the present
instruments. That will allow to probe the faintest end of the
luminosity function of the MSPs in GCs. If the current extrapolations
of this luminosity function (Ransom et al. 2005; Camilo \& Rasio 2005)
will turn out to be correct, an order of magnitude more MSPs could be
found in the core of the Galactic GCs, that have been missed by the
current surveys due to their relative faintness.  In this case,
$N^{\rm exp}_{\rm tot}\approx 1$ and SKA will be able to detect all of
this kind of systems. Therefore, a complete search for MSPs in the GCs
of the Milky Way with SKA will have the potentiality of testing the
hypothesis that IMBHs of order $100\,\msun$ are commonly hosted in
GCs.
  
The detection of one [IMBH,MSP] system will immediately give the
chance of measuring the mass of the IMBH from pulsar timing with at
least $1\%$ accuracy (Cordes et al. 2004). Even more interesting, the
presence of a very stable clock (like MSPs usually are) orbiting a
probably rotating $\sim 100\,\msun$ black hole makes this system a
potentially unique laboratory of relativistic physics. In fact, many
still elusive higher order relativistic effects depend on the spin and
on the quadrupole moment of the rotating black hole (Wex \& Kopeikin
1999) and the latter two quantities scale with the mass squared and
the mass cubed of the BH, respectively. Therefore, an [IMBH,MSP]
binary is a more promising target for studying the physics in the
surroundings of a BH (Kramer et al. 2004) than a binary comprising a
MSP and a stellar mass BH.

\section{Summary}

In this paper, we investigated the dynamical processes leading to the
capture of a MSP by an IMBH in the dense core of a GC.  We simulated
single-binary and binary-binary encounters between an IMBH and a MSP,
either single or with a WD companion.  The binary MSPs have masses and
orbital parameters chosen according to the distribution observed in a
sample of 23 GCs.  In order to account for all the possible
configurations of IMBHs hosted in GCs, we have considered the case of
a single IMBH, of an [IMBH,star] binary and of an [IMBH,BH] binary.
For each of these cases we derived the cross-section for the formation
of [IMBH,MSP] and [IMBH,WD] binaries, as well as the distribution of
the final semi-major axes and eccentricities of such newly formed
binaries.

The main outcomes from this study are:
\begin{itemize}
\item[$\bullet$] Dynamical encounters of a MSP with either single
  IMBHs or [IMBH,star] binaries promote the formation of [IMBH,MSP]
  binaries in $\sim 10\%$ and $\sim 1-5\%$ of the calculated
  interactions, respectively.  Similar rates were found for the
  formation of [IMBH,WD] binaries.  The final distributions of
  semi-major axes and eccentricities of the formed [IMBH,MSP] and
  [IMBH,WD] binaries are found to be in agreement with previous
  semi-analytical models (Pfahl 2005).
\item[$\bullet$] We found that the presence of a stellar mass BH,
  orbiting around the IMBH, strongly inhibits the formation of an
  [IMBH,MSP] binary. Only in a small minority of cases
  ($\sim{}0.2$\%), interactions between an [IMBH,BH] binary and a MSP
  can allow for the formation of a stable hierarchical triple, where
  the MSP occupies the external orbit. When the internal [IMBH,BH]
  binary merges due to orbital decay by gravitational waves emission,
  the triple evolves into a new [IMBH,MSP] binary.
\item[$\bullet$] The [IMBH,MSP] binaries are expected to form with
  very high eccentricities ($e\sim{}0.9$) and tight orbits ($\lsim 7$
  AU). This means that they can be important sources of gravitational
  waves, either in the in-spiral phase or in the final merging event.
\item[$\bullet$] Due to the aforementioned gravitational quadrupole
  radiation, the [IMBH,MSP] binaries are relatively short-lived,
  in-spiraling to coalescence in $\sim 10^8$ yr. This lifetime is
  significantly shorter than the estimated formation timescale of
  [IMBH,MSP] binaries which may be detectable with the present
  instrumentation.
\item[$\bullet$] If IMBHs of $\sim$ 100$\,\msun$ are commonly hosted in
  the Galactic GCs, next--generation radio telescopes, like SKA, will
  have the possibility of detecting at least one of these exotic
  binaries.
\end{itemize}

\section*{Acknowledgments}
We thank S.~Aarseth for enlightening discussions and for having kindly
provided us the code Chain. We thank the Referee for her/his critical
comments that allowed us to significantly improve the manuscript. MC and AP
acknowledge financial support from MURST, under the grant PRIN-2005024090.

\appendix

\section {Initial Conditions}

\subsection{Initial semi-major axis distribution}
We describe here in some detail how we generate the initial
distribution for the semi-major axis of the IMBH binaries.
  
\begin{figure}
\begin{center}
\includegraphics[width=7cm]{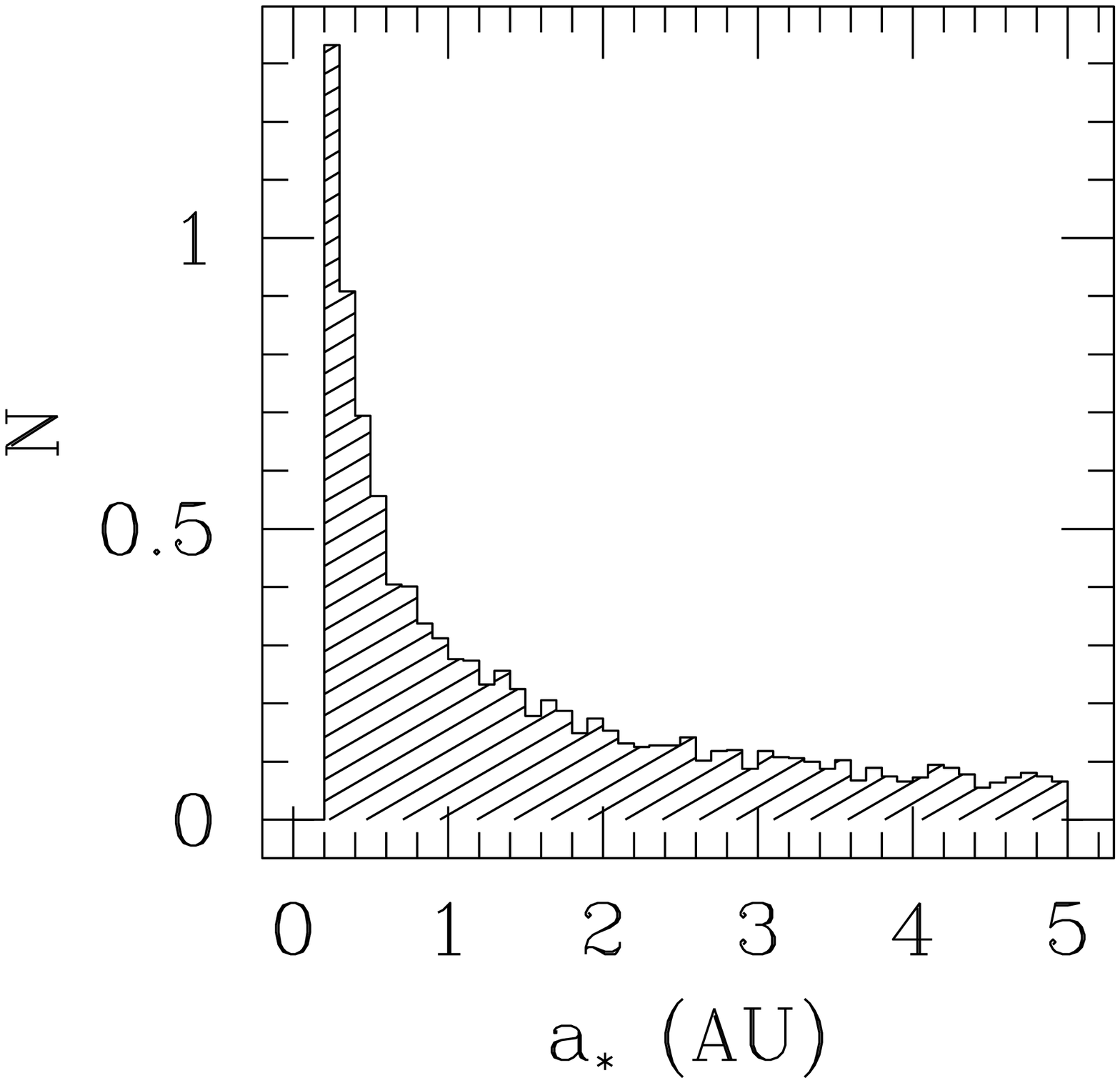} 
\includegraphics[width=7cm]{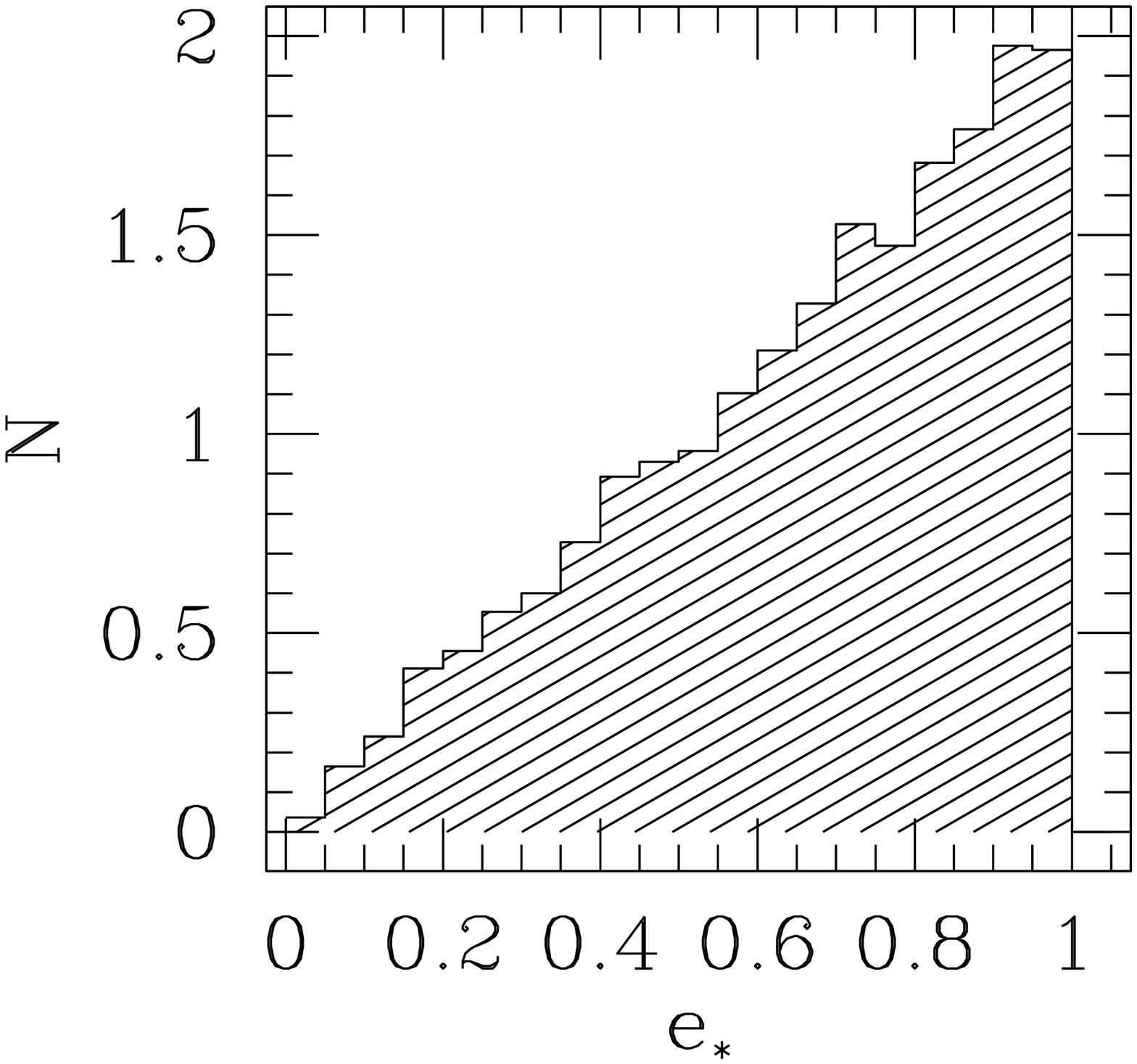}
\caption{
  Initial distribution of the semi-major axes (left panel) and
  eccentricities (right panel) of the initial states [IMBH,star]
for the $M_{\rm IMBH}=100\,\msun$.  }
\label{fig:A1}
\end{center}
\end{figure}

$\bullet$ [IMBH,star]: We have followed the analysis of Pfahl (2005)
who considers the tidal disruption of a stellar binary off an IMBH.
From considerations on energy conservation, the semi-major axes $a_*$
of the newly formed binary follows the relation 
\begin{center}
\begin{equation}
a_*=\frac{1}{2}a_{\rm b}
\frac{m_{\rm b}}{m_{\rm esc}}\left(\frac{M_{\rm IMBH}}{m_{\rm b}}
\right)^{2/3},
\end{equation}
\end{center}
\noindent
where $m_{\rm b}$ is the mass of the initial binary, $a_{\rm b}$ its
semi-major axis, and $m_{\rm esc}$ the mass of the escaping star (see
also the discussion in Section 3.2).  To reproduce the initial
distribution for $a_*$, we have considered a uniform distribution for
the mass ratio of the stellar binary $q\equiv {m_1}/{m_2}$ and a
distribution homogeneous in ${\rm log\left(a_b\right)}$ for the values
of the semi-major axes of the incoming binary in the range [0.01,10]
AU.  The upper and lower limits obtained are reported in Table~1.
Fig. \ref{fig:A1} shows the initial distributions of $a_*$ (left
panel) and $e_*$ (right panel). Note that the distribution of $a_*$ is
harder than that found in Blecha et al. (2006)\footnote{ We note that
  in their simulation, Blecha et al. consider stellar cluster
  considerably different from ours. Indeed, they study the formation
  of [IMBH,star] binaries in young clusters (their simulation stops
  after 100 Myr) with a correspondingly different population of
  stellar binaries. We argue that this can be the main cause of the
  difference in the distribution of $a_*$.}.  If the real distribution
would be less hard as in Blecha et al., our resulting [IMBH,MSP]
formation rates for the [IMBH,star] case should be considered as a
lower limit. Indeed a less bound initial companion to the IMBH would
be more easily ejected by the unstable triple interaction with the MSP
(see Section 3.2).

\noindent
$\bullet$ [IMBH,BH]: We expect that [IMBH,BH] binaries can form
dynamically in the core of a GC.  If the IMBH has been formed through
a succession of gravitational encounters with stellar-mass BHs, then
we expect some of these to be ejected in the outer region of the GC
and to sink back to the core by dynamical friction (Sigurdsson \&
Hernquist 1993).  The formation of the [IMBH,BH] binary can then be
the result of one of the following interactions:

\noindent
IMBH+[BH,star] $\rightarrow$ [IMBH,BH]+star;

\noindent
[IMBH,star]+[BH,star] $\rightarrow$ [IMBH,BH]+stars;

\noindent
[IMBH,star]+ BH  $\rightarrow$ [IMBH,BH]+star.

The [IMBH,BH] binary just formed is assumed to have a separation
comparable to the IMBH influence radius.  This is not our initial
condition for simulating the encounters with the [MSP,WD] systems,
since we have accounted for the intrinsic long term evolution of the
[IMBH,BH] binary parameters.  Accordingly, we have generated the
values of the initial [IMBH,BH] binary semi-major axis, (i.e.  the
values of $a_{\rm BH}$ from which we start the 3 or 4 body
simulations) from a distribution obtained sampling uniformly in time
when considering the evolution of $a_{\rm BH}$ due to the hardening
(i) off cluster stars, and (ii) by gravitational wave back-reaction.

In phase (i), denoted in the text as [IMBH,BH]$_{\rm h,*}$, the
evolution of $a_{\rm BH}$ is governed by the equation:
\begin{equation}\label{eq:aharde}
\frac{\mbox{d}a_{\rm BH}}{\mbox{d}t}=-\left(2\pi\xi\right)
\frac{G\langle \rho_*\rangle }{\langle v_*\rangle }a^2_{\rm BH}, 
\end{equation}
holding until $a_{\rm BH}=a_{\rm gw}(e_{\rm BH}=0.7)$ set by equation
(\ref{eq:agw}) (Hills 1975).  In equation (A2) we assumed fixed the
values of $\langle \rho_*\rangle =7\times 10^5\,\msun {\rm pc^{-3}}$
and $\langle v_* \rangle=10 \,{\rm km s^{-1}}$ inferred averaging over
the current GC sample described in Section 6.  A change in $\langle
\rho_*\rangle$ and $\langle v_*\rangle$ due to the internal evolution
of the GC should also change the $a_{\rm BH}$ distribution. In
particular, a lower value for the stellar density should enhance the
right end tail of the distribution. We argue that in this case the
formation of [IMBH,MSP] binaries could be enhanced.  Indeed, the
presence of an initial companion, bound to the IMBH on a less tight
orbit than that considered in our study, would be more easily ejected
by the MSP (see Section 3.2).

\begin{figure}
\begin{center}
\centerline{\psfig{figure=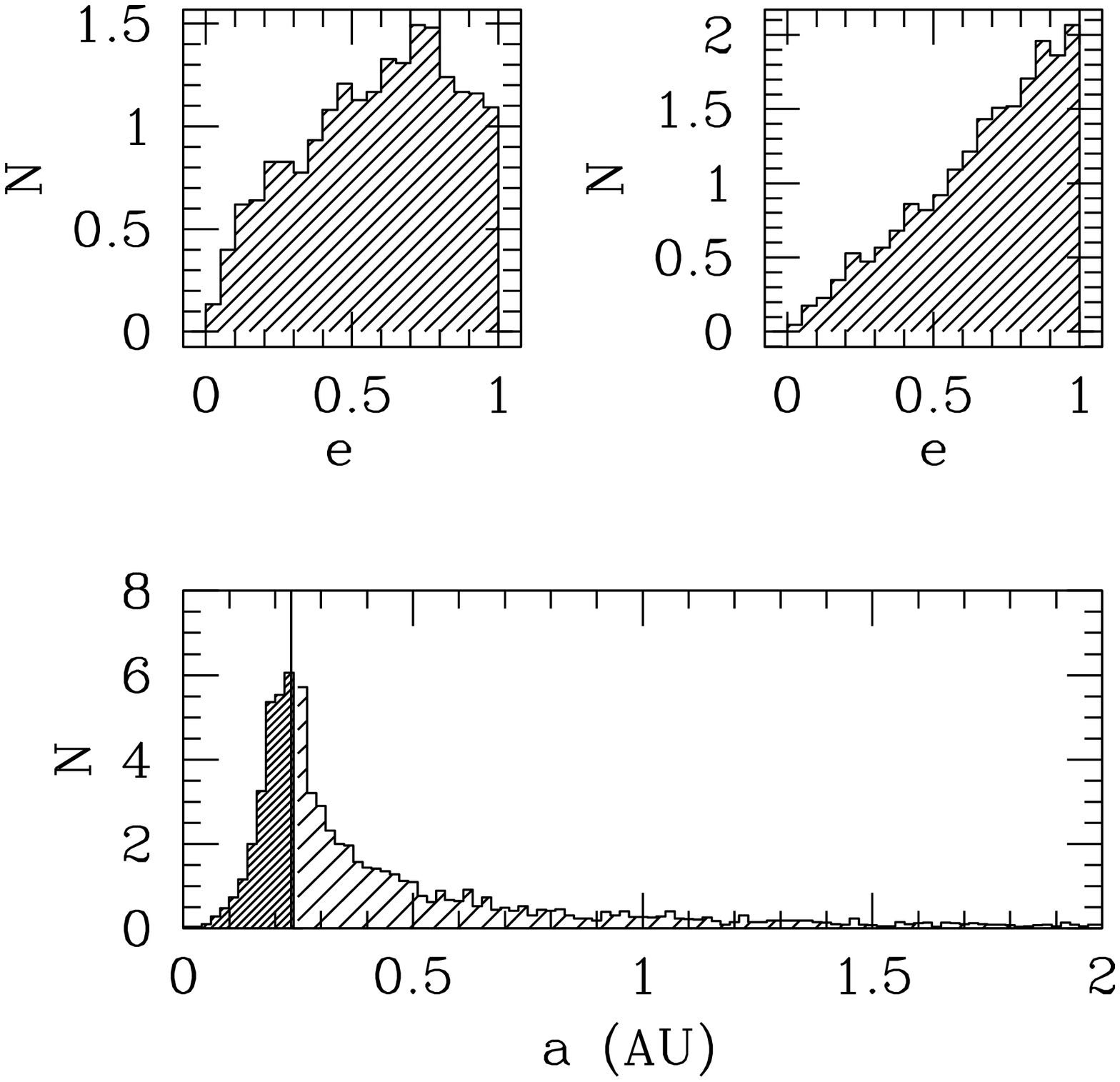,height=8.cm}}
\caption{ 
Upper panel: the
distribution for the initial eccentricity of the [IMBH,BH] binaries
are shown in the regimes of hardening off stars (right) and
gravitational waves (left), for $M_{\rm IMBH}=100\,\msun$.  
Lower panel: the distribution of the initial
semi-major axes for the same systems are shown: the solid vertical
line separates the gravitational wave regime (left) and the hardening
off stars regime (right).}
\label{fig:ain100}
\end{center}
\end{figure}

In phase (ii), the binary hardens by gravitational
waves back--reaction (phase denoted with [IMBH,BH]$_{\rm gw}$).  
The evolution of the orbital parameters are given by (Peters
1964):
\begin{center}
\begin{equation}\label{eqn:dagw}
\frac{\mbox{d}a_{\rm BH}}{\mbox{d}t}=-\frac{64}{5}\frac{G^3m_{\rm BH}
M_{\rm IMBH}\left(m_{\rm BH}+M_{\rm IMBH}\right)}{c^5a^3_{\rm BH}}f(e_{\rm BH})
\end{equation}
\end{center}
\begin{center}
\begin{equation}\label{eqn:degw}
\frac{\mbox{d}e_{\rm BH}}{\mbox{d}t}=-\frac{304}{15}\frac{G^3m_{\rm BH}
M_{\rm IMBH}\left(m_{\rm BH}+M_{\rm IMBH}\right)}{c^5a^4_{\rm BH}}g(e_{\rm BH})
\end{equation}
\end{center}
\noindent
where

\begin{center}
\begin{equation}
f(e_{\rm BH})=\left(1-e_{\rm BH}^2\right)^{-7/2}
\left(1+\frac{73}{24}e^2_{\rm BH}+\frac{37}{96}e^4_{\rm BH}\right)
\end{equation}
\end{center}

\begin{center}
\begin{equation}
g(e_{\rm BH})=\left(1-e^2_{\rm BH}\right)^{-5/2}
e_{\rm BH}\left(1+\frac{121}{304}e^2_{\rm BH}\right). 
\end{equation}
\end{center}
The above equations (A3-A6) are integrated with the initial condition:
$a_{\rm BH}$ =$a_{\rm gw}(e_{\rm BH})$ and a trial distribution for
$e_{\rm BH}$ that follows the thermal distribution.  Fig. A2 shows
the resulting distribution for $a_{\rm BH}$ and $e_{\rm BH}$ during
the two different regimes.  \footnote{Note that in phase (ii), the
  distribution should not be affected by any change in the structural
  parameters of the GC, depending only on the orbital parameters (see
  equations (A3) and (A4)).}

\subsection{The integration}

In this subsection we describe details on the integration of
three--body and four--body encounters with the codes Chain and FEBO.
For each run we divide the integration into two parts.  (I) We
consider the two binaries as point-like objects until their centers of
mass are at a distance larger than 50 times the semi-major axis of the
IMBH binary. (II) When this critical distance is reached, we start the
four- body integration. As a consequence, the time spent in the two
body approximation decreases as the semi-major axes of the IMBH binary
become wider.  Correspondingly, the overall integration time gets
longer the wider the semi-major axis of the IMBH binary is, and it
becomes prohibitively long for large values of $a_{\rm BH}$ (or $a_*$
for the [IMBH,star] binaries).  For this reason, we insert a cut-off
at 5 AU.  For wider systems, we expect that the available binding
energy of the IMBH binary is insufficient to unbind the [MSP,WD]
binary, so that the ionization of the binary can be mainly due to the
tidal effect of the massive IMBH.

\subsection{Impact parameters}

In this subsection we focus attention on the choice of the maximum
impact parameter $b_{\rm max}$ for a correct determination of the
cross section.  According to gravitational focusing, a point mass with
impact parameter $b$, moving at infinity with relative velocity
$v_{\infty}$, has pericenter
\begin{center}
\begin{equation}
p\sim {b^2 \, v^2_{\infty}
\over 2 GM_{\rm IMBH} }.
\end{equation}
\end{center}

\begin{figure}
\begin{center}
\centerline{\psfig{figure=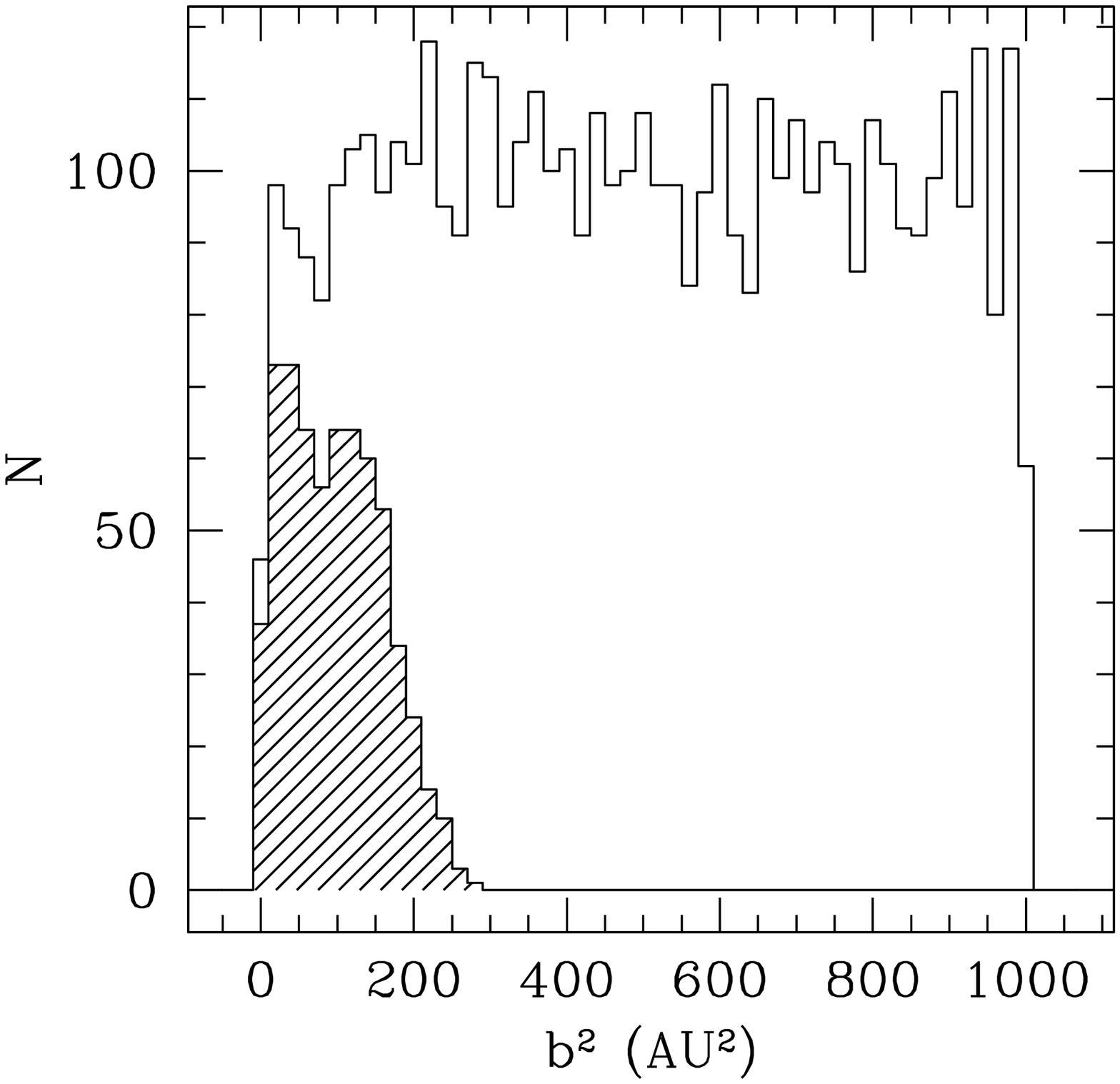,height=8cm}}
\caption{Distribution of impact parameters giving rise to the
  formation of a [IMBH,MSP] system (hatched histogram), compared to
  the initial (empty histogram) for the case of encounters of PSR-A
  like [MSP,WD] binaries off the single 100$\,\msun$ IMBH.  The hatched
  distribution drops to zero at 300 AU$^2$, while $b^2_{\rm max}=1000$ AU$^2$.
}
\label{fig:b100g}
\end{center}
\end{figure}

For the single IMBH, the maximum value of the periastron $p_{\rm max}$
is set at a few tidal radii $r_{\rm T}$; while for a binary IMBH, the
value of the maximum impact parameter for a non negligible energy
exchange is typically limited up to a value of the order of a few
semi-major axis of the binary IMBH, i.e. $p_{\rm max}\sim xa_{\rm BH}$
or $p_{\rm max}\sim xa_*$ (Hills 1983), where $x$ is close to 3 in all
cases.  In each run $b^2_{\rm max}$ is assigned using equation (A7).
In order to guarantee that we have accounted for all the impact
parameters leading to the formation of an [IMBH,MSP] (or [IMBH,WD])
binary, and to guarantee cross section convergence, we verified a
posteriori that the distribution of all relevant $b^2$ leading to the
desired end--states, drops to zero well before $b^2_{\rm max}.$ Fig.
\ref{fig:b100g} illustrates the case of the single $100 \,\msun$ IMBH
interacting with the PSR-A like [MSP,WD] binary.  The encounters
ending with the formation of [IMBH,MSP] binaries are the ones
represented in the hatched area. Clearly, the distribution drops to
zero well before $b^2_{\rm max}.$

In the
channels where either the tidal radius of the incoming
binary MSP, or the semi--major axis of the
IMBH binary vary from encounter to encounter (according to the
initial distributions described in Sections 2.2 for the binary IMBH,
and in Section 5 for the binary MSP), we allowed $b^2_{\rm max}$ to
vary accordingly, and defined a mean $\langle b^2_{\rm max}\rangle,$
obtained averaging over all choices of $r_{\rm T}$ and/or $a_*$
($a_{\rm BH}$). This average is used to compute the cross section in
equation (5).

\end{document}